\documentclass[iop]{emulateapj}

\usepackage{color}
\usepackage{amssymb, amsmath}

\usepackage[utf8]{inputenc}
\usepackage{lipsum}
\usepackage{pifont}
\usepackage{bm}
\usepackage{ulem}
\usepackage{balance}
\usepackage{multirow}
\usepackage{mathtools}
\usepackage{enumitem}
\usepackage{booktabs}
\usepackage{appendix}
\usepackage{listings}

\shorttitle{Neural networks as optimal estimators to marginalize over baryonic effects}
\shortauthors{Francisco Villaescusa-Navarro et al.}

\usepackage[usenames,dvipsnames]{xcolor}
\usepackage{hyperref}
\hypersetup{
    colorlinks = true,
    citecolor = {MidnightBlue},
    linkcolor = {BrickRed},
    urlcolor = {blue}
}

\newcommand{\be}{\begin{equation}}
\newcommand{\ee}{\end{equation}}
\newcommand{\ba}{\begin{eqnarray}}
\newcommand{\ea}{\end{eqnarray}}

\usepackage{soul}
\definecolor{nicegreen}{HTML}{2CA02C}
\definecolor{orange}{rgb}{1,0.5,0}

\begin{document}

\title{Neural networks as optimal estimators to marginalize over baryonic effects}

\author{Francisco Villaescusa-Navarro$^{1,\dagger}$, Benjamin D. Wandelt$^{2,3,4}$, Daniel Angl\'es-Alc\'azar$^{5,4}$, 
Shy Genel$^{4,6}$, Jose Manuel Zorrilla Mantilla$^{1}$, Shirley Ho$^{4,1}$, David N. Spergel$^{4,1}$}
\affil{$^1$ Department of Astrophysical Sciences, Princeton University, Peyton Hall, Princeton, NJ, 08544, USA}
\affil{$^2$ Institut d'Astrophysique de Paris, 98bis Boulevard Arago, 75014 Paris, France}
\affil{$^3$ Sorbonne Universites, Institut Lagrange de Paris, 98 bis Boulevard Arago, 75014 Paris, France}
\affil{$^4$ Center for Computational Astrophysics, Flatiron Institute, 162 5th Avenue, New York, NY, 10010, USA}
\affil{$^5$ Department of Physics, University of Connecticut, 196 Auditorium Road, U-3046, Storrs, CT, 06269, USA}
\affil{$^6$ Columbia Astrophysics Laboratory, Columbia University, 550 West 120th Street, New York, NY, 10027, USA}
\altaffiltext{$\dagger$}{villaescusa.francisco@gmail.com}

\begin{abstract}
Many different studies have shown that a wealth of cosmological information resides on small, non-linear scales. Unfortunately, there are two challenges to overcome to utilize that information. First, we do not know the optimal estimator that will allow us to retrieve the maximum information. Second, baryonic effects impact that regime significantly and in a poorly understood manner. Ideally, we would like to use an estimator that extracts the maximum cosmological information while marginalizing over baryonic effects. In this work we show that neural networks can achieve that. We made use of data where the maximum amount of cosmological information is known: power spectra and 2D Gaussian density fields. We also contaminate the data with simplified baryonic effects and train neural networks to predict the value of the cosmological parameters. For this data, we show that neural networks can 1) extract the maximum available cosmological information, 2) marginalize over baryonic effects, and 3) extract cosmological information that is buried in the regime dominated by baryonic physics. We also show that neural networks learn the priors of the data they are trained on. We conclude that a promising strategy to maximize the scientific return of cosmological experiments is to train neural networks on state-of-the-art numerical simulations with different strengths and implementations of baryonic effects. 
\end{abstract}


\section{Introduction}
\label{sec:introduction}

Cosmology is becoming a precise and accurate branch of physics. The $\Lambda{\rm CDM}$ model is now well-established, and accurately explains a large variety of cosmological observations. This model describes how the large-scale structure of the Universe originates from primordial quantum fluctuations in the very early Universe through amplification by non-linear gravitational evolution.

The $\Lambda$CDM model contains a set of parameters describing fundamental physical quantities, such as the energy fraction in dark matter and dark energy, the geometry and expansion rate of the Universe, and the sum of neutrino masses. One of the most important goals in modern cosmology is to determine the value of those cosmological parameters with the highest accuracy. The motivation for doing so is improving our knowledge on the fundamental constituents and laws of the Universe.

In order to constrain the value of the cosmological parameters, observational data are collected and summary statistics are computed from them. Next, predictions from theory are made for these summary statistics as a function of the value of the cosmological parameters. Finally, data is confronted with theory and bounds on the parameters are deduced.

It has been recently shown that much tighter constraints on the value of the cosmological parameters can be established by extracting the information embedded on small, non-linear scales \citep{Quijote, Cora_19, Massara_19, Chang_19, Arka_2020, Dai_2020, Friedrich_19, Dai_2019, Krause_2017}. This motivates the usage of these scales in order to maximize the scientific return of cosmological surveys. Unfortunately, two major theoretical obstacles appear in this regime. First, it is unknown what statistic will allow extracting the maximum information from non-linear scales\footnote{We note that in the case of Gaussian density fields, the power spectrum (or the two-point correlation function) is the statistic that will completely characterize the properties of those fields. However, most cosmological surveys observe non-Gaussian density fields.}. Second, poorly understood baryonic effects such as supernova and active galactic nuclei (AGN) feedback are believed to significantly affect the distribution and properties of both dark and baryonic matter on these scales. We will use the term \textit{baryonic effects} when referring to these processes. 

Ideally, we would like to use a summary statistic that allow us to extract the maximum information from the entire field (e.g. galaxy number density field or 21cm field), while marginalizing over baryonic effects at the same time. The purpose of this paper is to show that neural networks can achieve these two goals. Furthermore, we will show that neural networks can extract cosmological information buried in the regime dominated by baryonic effects.

To demonstrate this, we create two different types of toy mock data: 1) power spectra, representing a summary statistic, and 2) 2D Gaussian density fields. In both cases, the maximum cosmological information embedded into the data is known a priori. We then train neural networks to predict the value of the cosmological parameters from these data. Next, we use a simple prescription to mimic the effects of baryons on the data and repeat the above exercise. In both cases, we compare the constraints from the network against the theoretical floor, showing that neural networks can learn optimal unbiased estimators that extract all the available cosmological information from the data. The analysis on the two types of datasets is almost identical in order to show the robustness of our conclusions. 

We emphasize that we made use of these simplistic data sets since their theoretical information floor, as well as the optimal estimator, is well-known. However, the neural networks do not know anything about the structure of the data, that they learn by looking at the examples. Furthermore, in the case of baryonic effects, we never tell the network the scale where these effects show up, so the network has to learn that as well.

This paper is organized as follows. In Section \ref{sec:Toy_model_I} we present the data from the first toy model, the power spectra, and compare the performance of the traditional maximum likelihood estimator against neural networks. Then in Section \ref{sec:Toy_model_II} we describe the data from our second toy model, the 2D Gaussian density fields, and carry out the corresponding analysis with neural networks. Finally, we summarize the main results of this work in Section \ref{sec:Conclusions}.

\section{Toy model I: power spectrum}
\label{sec:Toy_model_I}

In this section we first explain how we generate data from our first toy model: mock power spectra. We then train neural networks with power spectra that may, or may not, be affected by baryonic effects. Next, we compare the results we obtain using the maximum likelihood estimate against those from the neural networks.

\subsection{Data}
\label{subsec:Pk}

\begin{figure}
\centering
\includegraphics[width=0.49\textwidth]{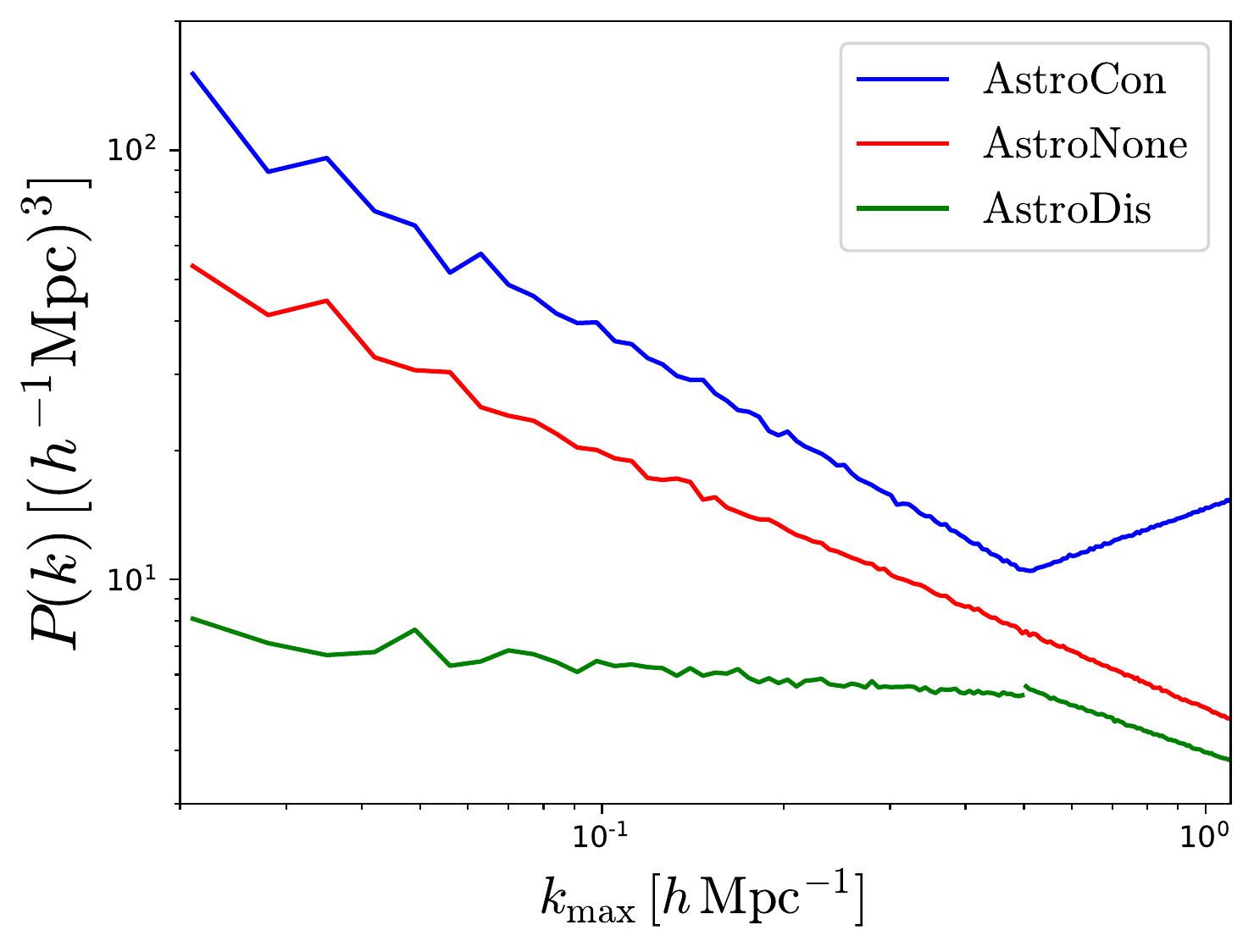}
\caption{Data examples from the first toy model: mock power spectra. The data consist of three different sets. AstroNone contains power spectra that are not affected by baryonic effects, hence its power spectra are simply power laws $P(k)=Ak^B$. AstroCon simulates the effect of baryons by implementing a different power law, $P(k)=Ck^D$, on scales $k>k_{\rm pivot}$. This model assumes that the power spectrum is continuous. AstroDis is identical to AstroCon, but does not require the power spectrum to be continuous. Cosmic variance, for a volume of $\simeq1~(h^{-1}{\rm Gpc})^3$, is added to the underlying power spectra on all scales.}
\label{fig:Toy_model_Pk}
\vspace{0.3cm}
\end{figure}

\begin{figure*}
\centering
\includegraphics[width=0.99\textwidth]{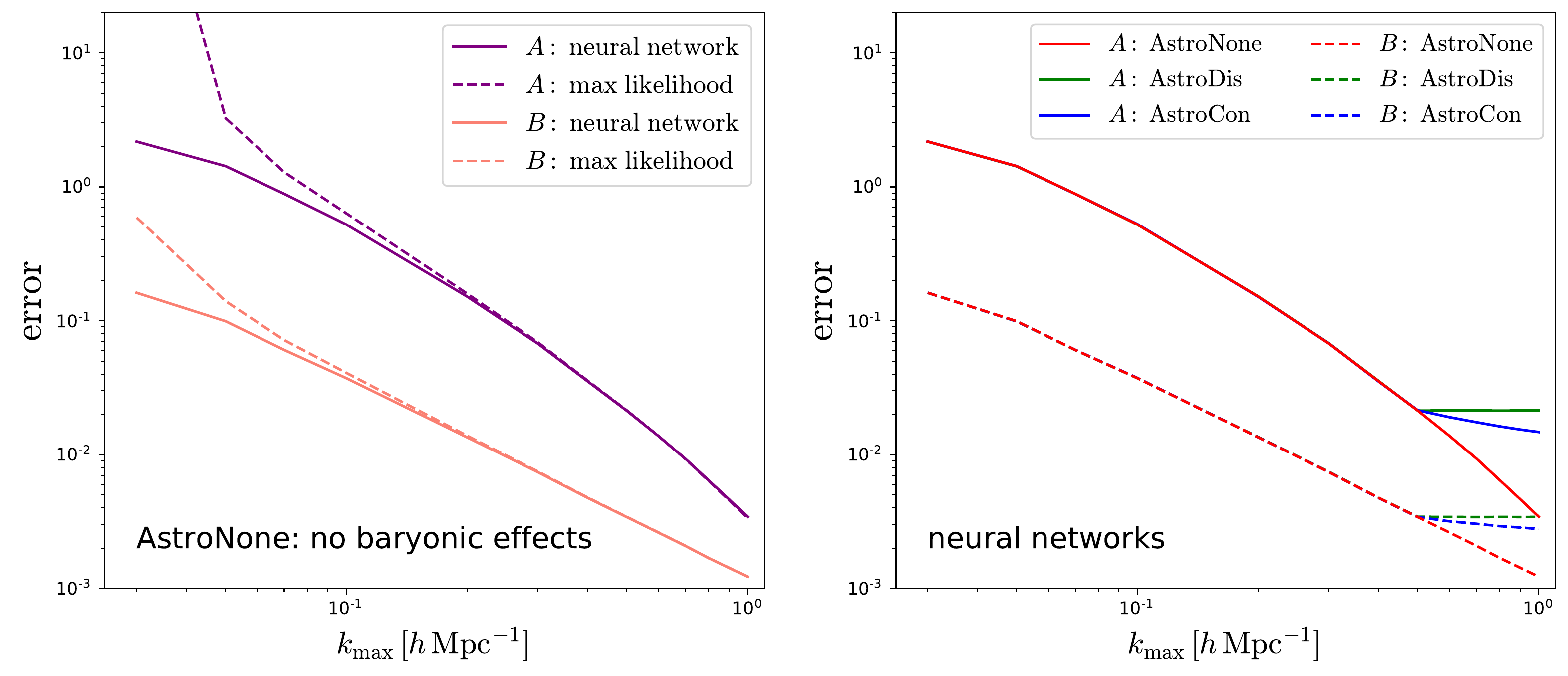}
\caption{\textbf{Left:} We train neural networks to predict the value of the cosmological parameters $A$ and $B$ from power spectra of the AstroNone set down to $k_{\rm max}$. The solid lines show the mean square error on $A$ and $B$ as a function of $k_{\rm max}$. We also determine the value of the cosmological parameters by maximizing the value of the likelihood function, and show the corresponding errors with dashed lines. We find that the neural network always performs equally, or better, than the maximum likelihood method, showing that neural networks can find the optimal solution to extract the cosmological information. \textbf{Right:} We train neural networks to predict the value of the cosmological parameters from power spectra of the AstroNone (red), AstroDis (green), and AstroCon (blue) sets, setting $k_{\rm pivot}=0.5~h{\rm Mpc}^{-1}$. We find that for $k\geqslant k_{\rm pivot}$ constraints on $A$ and $B$ from the AstroDis flattens out, pointing out that the network has learned the scale where baryonic effects show up, and it has learned to marginalize over scales smaller than that. For data from the AstroCon set, we find that going to higher $k_{\rm max}$ helps reducing the error on the cosmological parameters. This happens because the network has learned to extract the cosmological information that is buried in the regime dominated by baryonic effects (see text for further details).}
\label{fig:Toy_model}
\vspace{0.8cm}
\end{figure*}

The data from this toy model consist of simple power laws representing mock power spectra
\begin{equation}
P(k) = Ak^B~,
\end{equation}
where $A$ and $B$ are the \textit{cosmological parameters}. Our goal is to train neural networks to predict the value of $A$ and $B$ from measurements of the amplitude of the power spectrum in different $k-$bins. We use this very simple model, instead of more realistic power spectra from Boltzmann codes \citep{CAMB, CLASS} to keep things as simple and interpretable as possible.
\begin{table}
\begin{center}
\renewcommand{\arraystretch}{0.4}
\resizebox{0.49\textwidth}{!}{\begin{tabular}{| c | c | c | c |}
\hline
& \multicolumn{3}{|c|}{Dataset} \\[0.5ex]
\hline
\multirow{2}{*}{} & \multirow{2}{*}{AstroNone} & \multirow{2}{*}{AstroCon} & \multirow{2}{*}{AstroDis} \\
& & & \\
\hline
\multirow{2}{*}{$P(k)$} &  \multirow{2}{*}{$Ak^B$} & $Ak^B$ if $k\leq k_{\rm pivot}$ & $Ak^B$ if $k\leq k_{\rm pivot}$ \\
 & & $Ck^D$ if $k>k_{\rm pivot}$ & $Ck^D$ if $k>k_{\rm pivot}$\\
\hline
\hline
\multirow{2}{*}{A} &  \multirow{2}{*}{[0.1 , 10.0]} &  \multirow{2}{*}{[0.1 , 10.0]} &  \multirow{2}{*}{[0.1 , 10.0]} \\
 & & &\\
\hline
\multirow{2}{*}{B} &  \multirow{2}{*}{[-1 , 0]} &  \multirow{2}{*}{[-1 , 0]} &  \multirow{2}{*}{[-1 , 0]} \\
 & & &\\
 \hline
 \multirow{2}{*}{C} &  \multirow{2}{*}{-} &  \multirow{2}{*}{$Ak_{\rm pivot}^{B-D}$} &  \multirow{2}{*}{$Ak_{\rm pivot}^{B-D}\times$[0.5 , 1.5]}\\
 & & &\\[0.8ex]
 \hline
 \multirow{2}{*}{D} &  \multirow{2}{*}{-} &  \multirow{2}{*}{[-0.5 , 0.5]} &  \multirow{2}{*}{[-0.5 , 0.5]} \\
 & & &\\[0.5ex]
\hline
\end{tabular}}
\end{center}
\caption{Summary of the properties of the dataset of our first toy model: power spectra. We consider three different sets; AstroNone, AstroCon, and AstroDis. The functional form of the power spectra is given in the $P(k)$ row, while the range in which the cosmological ($A$ and $B$) and astrophysical ($C$ and $D$) parameters is varied is shown below.}
\label{table:Pk_models}
\end{table}
We consider three different datasets, depending on the way baryonic effects are modelled:
\begin{itemize}
\item \textbf{AstroNone}. This set assumes that there are no baryonic effects. Thus, the form of the power spectrum is just given by  $P(k)=Ak^B$.

\item \textbf{AstroDis}. This set incorporates baryonic effects, implemented as follows. On large scales, the power spectrum is not affected by baryons, and therefore, it just follows the power law $P(k)=Ak^B$. On scales $k>k_{\rm pivot}$, baryons affect the power spectrum, inducing a different power law\footnote{We note that baryonic effects are neither strictly multiplicative not strictly additive, so we simply represent them as a new power law below a pivot scale in our toy model. } $P(k)=Ck^D$.

\item \textbf{AstroCon}. This set implements baryonic effects in the same way as AstroDis, with the only difference being that here the power spectrum is required to be continuous at $k_{\rm pivot}$, implying that $Ak_{\rm pivot}^B=Ck_{\rm pivot}^D$.

\end{itemize}
In all datasets, the values of $A$ and $B$ are drawn from uniform distributions from 0.1 to 10, and -1.0 to 0, respectively. For AstroCon and AstroDis, $D$ is taken randomly between -0.5 and +0.5, following a uniform distribution. For AstroCon, the value of $C$ is fixed to $\bar{C}=Ak_{\rm pivot}^{B-D}$, while for AstroDis, $C$ is sampled from a uniform distribution between $0.5 \bar{C}$ and $1.5\bar{C}$.

Once the amplitude and shape of the power spectrum is known, we consider a set of $k-$bins $k\in[3k_F, 4k_F, 5k_F,...k_{\rm max}]$, where $k_F$ is the fundamental frequency and $k_{\rm max}$ represents the smallest scale we consider. We take $k_F$ to be $7\times10^{-3}~h{\rm Mpc}^{-1}$, corresponding to a volume of $\simeq~1~(h^{-1}{\rm Gpc})^3$. We start from $3\times k_F$, instead of $k_F$, to avoid negative values on the power spectrum when adding cosmic variance (see below).

For each realization, we generate a \textit{measured} power spectrum with no noise but with cosmic variance as follows. For each $k-$bin, $k_i$, we draw the amplitude from a Gaussian distribution with mean $\mu_i=P(k_i)$, and variance $\sigma_i^2=2P^2(k_i)/N_{k_i}$, where $N_{k_i}=4\pi k_i^2k_F/k_F^3$ is the number of modes in the considered $k-$bin. For simplicity, we assume that the different scales are independent, thus, the covariance matrix is diagonal.

Fig.~\ref{fig:Toy_model_Pk} shows examples of power spectra from the AstroNone, AstroCon, and AstroDis datsets. Table \ref{table:Pk_models} summarizes the characteristics of the different sets. Generating these mock power spectra is so fast that we train the networks with data generated on the fly.

\subsection{Neural networks}
\label{subsec:NN1}

We train several simple neural networks to predict the value of the cosmological parameters, $A$ and $B$, from measurements of the amplitude of the power spectrum in different $k-$bins down to a maximum $k$ of $k_{\rm max}$. Each network is trained for a different value of $k_{\rm max}$.

Our architecture consists of four fully connected hidden layers, with leaky ReLU activation functions. The hidden fully connected layers have 60 neurons each. We use the Adam optimizer with beta parameters of 0.9 and 0.999. We use a learning rate of $10^{-4}$ and a batch size of 128. When the loss flattens out, we decrease the learning rate by a factor between 5 and 10. We repeat this procedure until we observe no further improvement. We emphasize that since producing these power spectra is computationally very cheap, we can generate as many of them as desired. Thus, overfitting is not a concern in our model. Depending on the value of $k_{\rm max}$ and the data set, we use between 10 million and 1 billion power spectra to train the networks. 

We first train the networks on power spectra from the AstroNone dataset, which do not incorporate baryonic effects. Once the model is trained, we test its accuracy using a set of $N=100,000$ power spectra. The left panel of Fig.~\ref{fig:Toy_model} shows with solid lines the error on $A$ and $B$ as a function of $k_{\rm max}$. The error is defined as the mean square error between the prediction of the neural network and the true value
\begin{equation}
{\rm error} = \frac{1}{N}\sum_{i=1}^N \left( X_{{\rm NN},i} - X_{{\rm true},i}\right)^2~,
\label{eq:error_NN}
\end{equation}
where $X$ can be either $A$ or $B$. $X_{{\rm NN},i}$ and $X_{{\rm true},i}$ are the prediction of the neural network and the true value, respectively, of the parameters for the $i$-th power spectrum.

As expected, the error on the parameters shrinks as $k_{\rm max}$ increases: more modes are available and their cosmological information is extracted by the network. 

\subsection{Optimal estimator}
\label{subsec:Theory_bounds}

Since our data is so simple, we can write down its exact likelihood. Taken into account that the amplitude of the power spectrum in each $k-$bin follows a Gaussian distribution, and that there is no correlation between different scales, the likelihood for a power spectrum with $m$ $k-$bins will be given by
\begin{widetext}
\begin{equation}
\mathcal{L}=\frac{1}{\sqrt{(2\pi)^m\delta P(k_1)\delta P(k_2)\cdot\cdot\cdot \delta P(k_m)}}\exp\left(-\frac{1}{2}\sum_{i=1}^m\left(\frac{P(k_i)-Ak_i^B}{\delta P(k_i)}\right)^2\right)~.
\label{Eq:likelihood}
\end{equation}
\end{widetext}
where $\delta P(k_i)$ is the error on the amplitude of the power spectrum of bin $i$. We note that in the above function, the errors on the power spectrum are computed as $\delta^2P(k_i)=2(Ak_i^B)^2/N_{k_i}$, namely based on the amplitude of the power spectrum before adding cosmic variance. 

For a given power spectrum, we can determine the value of $A$ and $B$ that maximizes the likelihood. Given a set of predictions for these parameters from multiple power spectra, we can quantify the \textit{error} on the parameters inherent to this method by using Eq.~\ref{eq:error_NN}; we just replace the prediction of the neural network by the maximum likelihood estimate.

The dashed lines in the left panel of Fig.~\ref{fig:Toy_model} show the results using the maximum likelihood estimate as a function of $k_{\rm max}$. For large values of $k_{\rm max}$, the accuracy on the parameters reached by the neural network equals that of the maximum likelihood estimate. This shows how our neural network is behaving as the optimal estimator with the lowest variance needed to extract all the cosmological information embedded in the power spectra.

Interestingly, we find that for low values of $k_{\rm max}$, the neural network estimator has a much lower variance than the one of the maximum likelihood method; for $k_{\rm max}=0.05~h{\rm Mpc}^{-1}$, the neural network predicts the value of $A$ and $B$ with an accuracy $\sim2.3\times$ and $\sim1.4\times$ better than the maximum-likelihood method. The reason for this, as we shall see below, is that the neural network \textit{learns} the priors on the distribution of the parameters.

\begin{figure}
\centering
\includegraphics[width=0.49\textwidth]{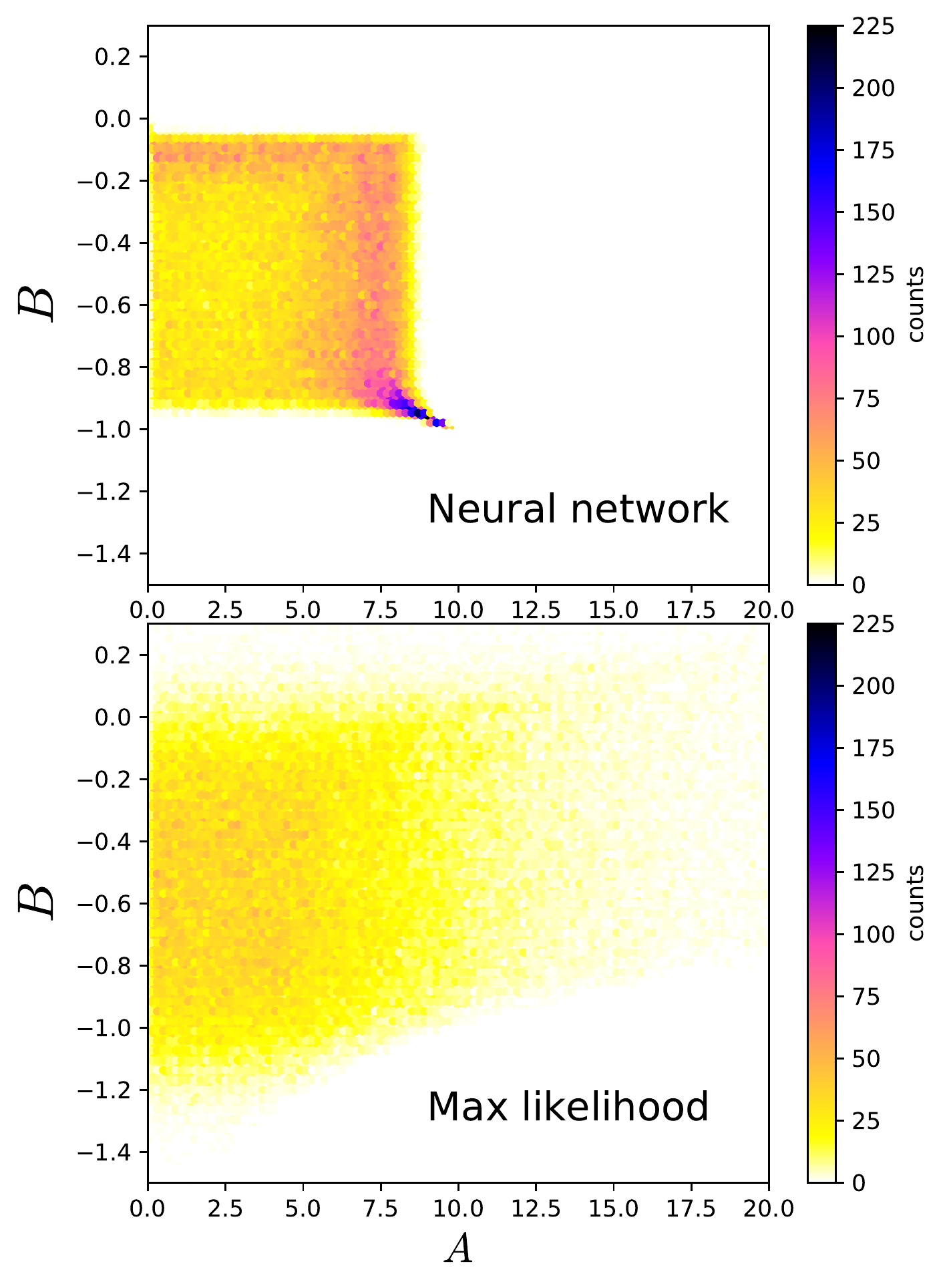}
\caption{We take 100,000 power spectra from the AstroNone set with $k_{\rm max}=0.05~h{\rm Mpc}^{-1}$, and predict the values of the cosmological parameters $A$ and $B$ for each using the neural network (top panel) and the maximum-likelihood (bottom panel) methods. Cells are color-coded according to the number of points within each cell. The neural network always predicts values of $A$ and $B$ within the range [0.1,10] and [-1,0], while the least squares method output values in a much wider range: $A\in[0,425]$, $B\in[-1.7,0.8]$. This shows how the neural network has learned the priors on the distribution of the parameters and never predicts values outside that range. This is the reason why the variance of the neural network is lower than the one of the maximum likelihood method.}
\label{fig:AB_values}
\end{figure}

\begin{figure*}
\centering
\includegraphics[width=0.49\textwidth]{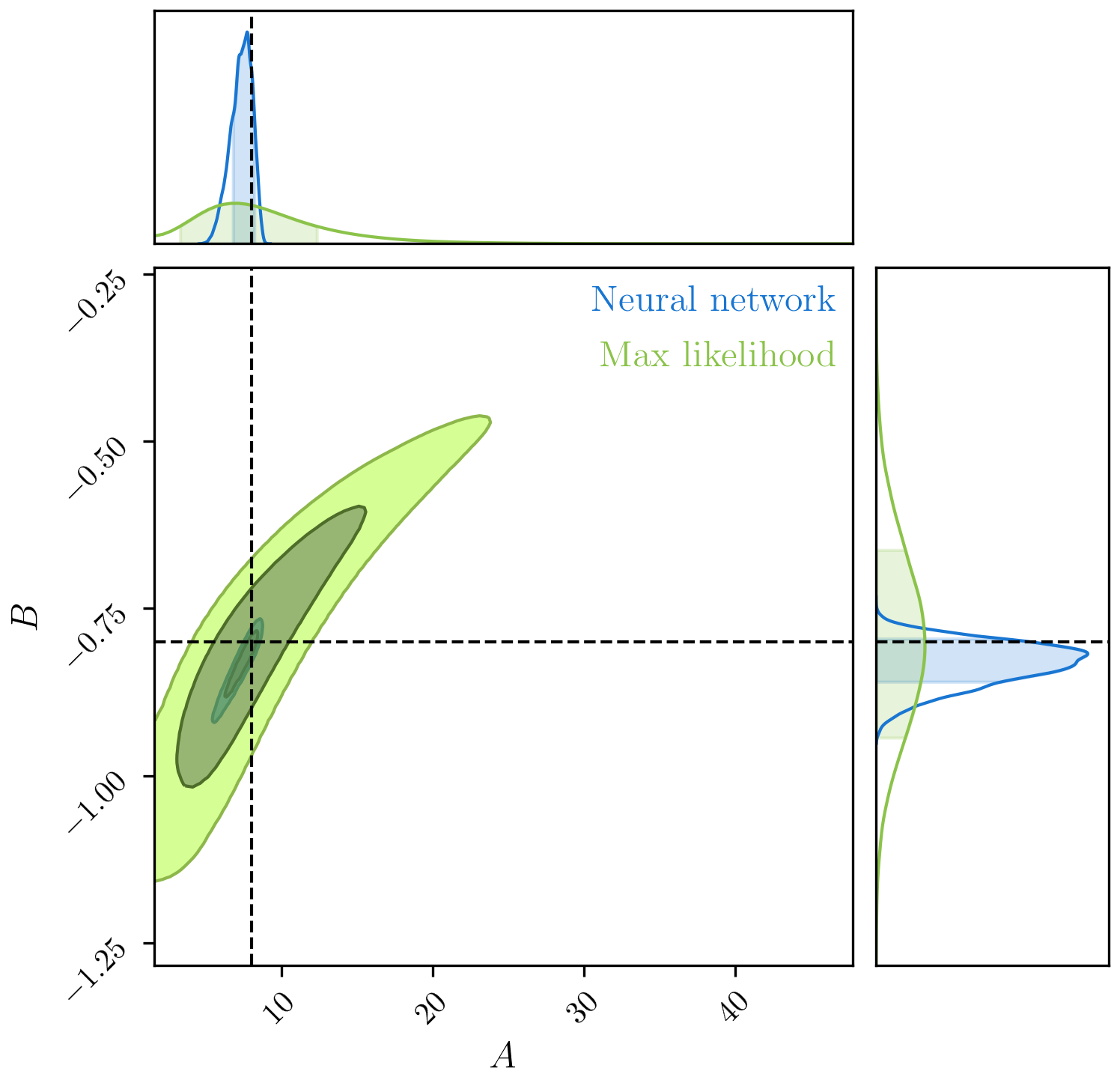}
\includegraphics[width=0.49\textwidth]{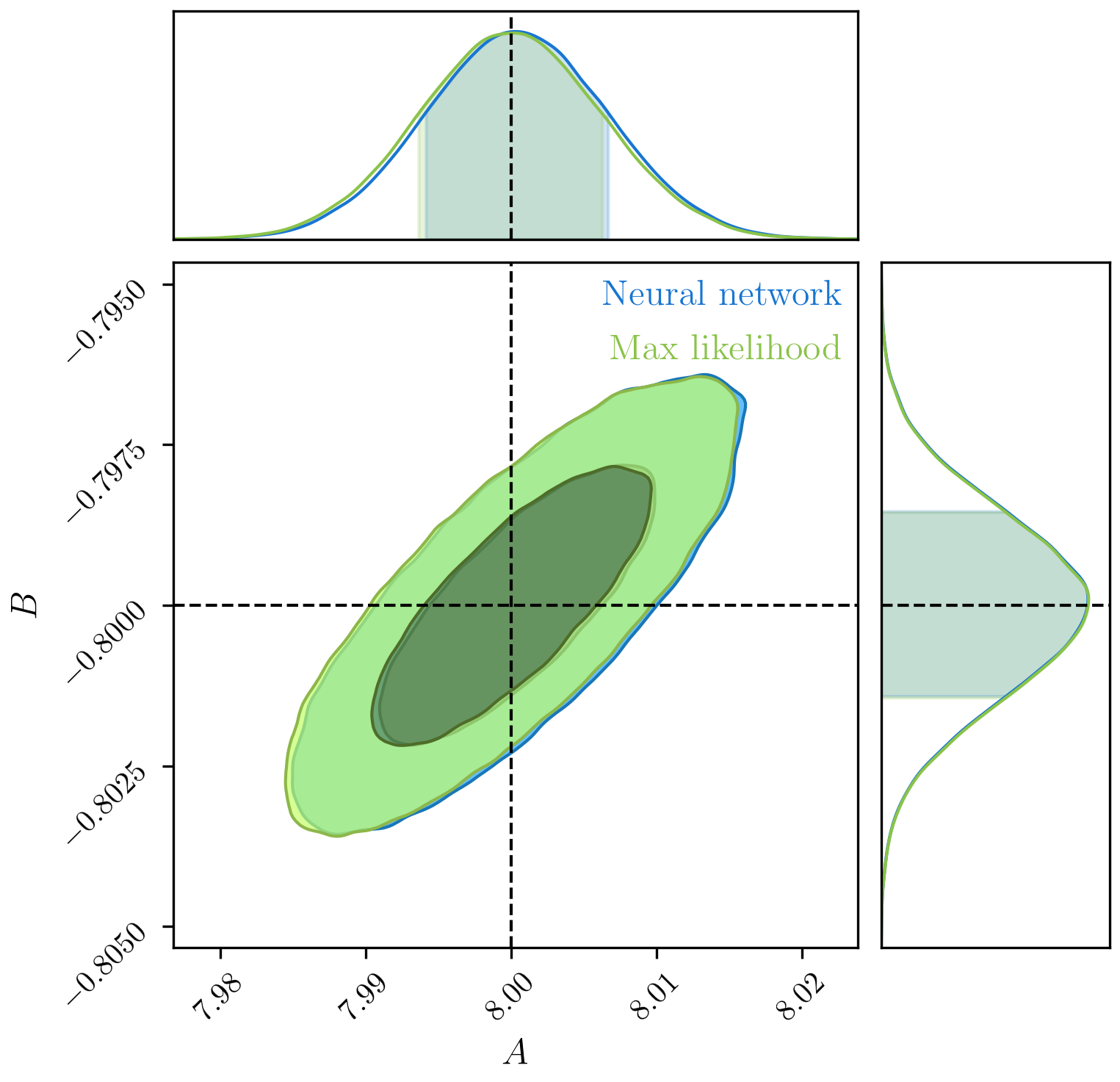}
\caption{We have generated 100,000 power spectra
with no baryonic effects (AstroNone set) but the same value of $A=8$ and $B=-0.8$. We input this data into the networks trained for $k_{\rm max}=0.05~h{\rm Mpc}^{-1}$ (left) and $k_{\rm max}=0.90~h{\rm Mpc}^{-1}$ (right). The panels show the distribution of the parameters output by the network, together with their projected 1D distributions. For $k_{\rm max}=0.90~h{\rm Mpc}^{-1}$ we see that the network and the maximum likelihood approaches produce almost identical results. On the other hand, for $k_{\rm max}=0.05~h{\rm Mpc}^{-1}$ we see that the distribution of the parameters is much larger for the maximum likelihood than for the network. From the 1D distributions we can see that the output of the network is affected by the priors the network was trained on: for instance, the network never saw a power spectrum with $A$ higher than 10, so the network never predicts values of $A$ above that. In this case, the network is computing a biased estimator of the parameters.}
\label{fig:Ellipses}
\end{figure*}

To show this, we take 100,000 power spectra from the AstroNone dataset and input them into the network trained for $k_{\rm max}=0.05~h{\rm Mpc}^{-1}$. For each power spectrum we estimate the value of $A$ and $B$ using the output of both the neural network and the maximum likelihood method. We show the resulting distributions of $A$ and $B$ in Fig.~\ref{fig:AB_values}. We find that the neural network always predicts the value of $A$ and $B$ to be within [0.1 , 10] and [-1 , 0], respectively; those values correspond to the range of variation of $A$ and $B$ in the training set. On the other hand, the maximum likelihood method predicts values well outside that range. This shows how the neural network has learned the priors on the distribution of the cosmological parameters, explaining why the variance of the network estimate is lower than the one of the maximum likelihood method. 

One important thing to note is that the distribution of the parameters $A$ and $B$ predicted by the network is not uniform within the priors, but presents a rich and complex structure. We investigate in more detail the structure of the space, together with the role played by the priors on the maximum likelihood estimate in the Appendix \ref{sec:priors}.

In order to better understand the effect of priors on the network, we have generated 100,000 power spectra from the AstroNone set that have the same value of the cosmological parameters: $A=8$ and $B=-0.8$. We have input these maps into the networks trained for $k_{\rm max}=0.9~h{\rm Mpc}^{-1}$ and $k_{\rm max}=0.05~h{\rm Mpc}^{-1}$, as well as to the maximum likelihood pipeline. We show the distribution of the parameters for these configurations in Fig. \ref{fig:Ellipses}. For $k_{\rm max}=0.90~h{\rm Mpc}^{-1}$, we can see that the network has found an unbiased estimator of the parameters, whose distribution matches almost perfectly the one from the maximum likelihood. On the other hand, for $k_{\rm max}=0.05~h{\rm Mpc}^{-1}$, the distributions of the parameters are very different. In the case of the neural network, the parameters are concentrated into a smaller region that is bounded by the priors, while the maximum likelihood expands a much broader area. In this case, the network is behaving as a biased estimator of the parameters. We however note that the variance of the neural network estimator is much smaller than the one of the maximum likelihood. 

In the Appendix \ref{sec:posterior_mean} we show that by construction, the network learns to approximate the posterior mean, which accounts for the prior. We shall see below, that similar behaviour is observed in the case of 2D Gaussian density fields.

\subsection{Baryonic effects}
\label{subsec:baryonic_effects_I}

We now investigate how accurately the network can predict the value of the cosmological parameters given power spectra that are affected by baryonic effects. We train a neural network using power spectra from the AstroDis set setting $k_{\rm pivot}=0.5\,h{\rm Mpc}^{-1}$. The green lines on the right panel of Fig.~\ref{fig:Toy_model} show the error on the parameters achieved by the neural network. 

For $k\leqslant k_{\rm pivot}$ the network can constrain the value of the cosmological parameters with the same accuracy as the network trained using power spectra from the AstroNone set. This is expected, since baryonic effects only appear on scales $k> k_{\rm pivot}$. For $k\geqslant k_{\rm pivot}$ we find that constraints saturate, i.e. no improvement on the cosmological parameters can be achieved by going to smaller scales. This is expected, because on those scales, the power spectrum follows a power law $P(k)=Ck^D$, where both $C$ and $D$ are not related to the cosmological parameters\footnote{This is not strictly true as the value of $C$ is drawn from an uniform distribution $\mathcal{U}[0.5-1.5]\bar{C}$, where $\bar{C}=Ak_{\rm pivot}^{B-D}$. However, in practice, that range is large enough to consider that $C$ is independent of $A$ and $B$.} $A$ and $B$.

This shows how neural networks can learn to marginalize on scales where baryonic effects dominate and no cosmological information is available. We emphasize that we did not input any information to the network with respect to $k_{\rm pivot}$. The network has learned that scale from the examples it has been trained on. 

By repeating the above exercise with the maximum likelihood method we find that in the cases where $k_{\rm max}\leq k_{\rm pivot}$ we obtain the same results as in the AstroNone case, as expected. On the other hand, for values of $k_{\rm max}> k_{\rm pivot}$, the error on the parameters rises dramatically. This is expected, as in this case, the likelihood function we wrote in Eq. \ref{Eq:likelihood} does not describe the baryonic effects imprinted in these power spectra. In order to do this analysis in a proper way with the maximum likelihood method, we would need to specify the value of $k_{\rm pivot}$ and truncate the likelihood function at that scale. These things, on the other hand, are automatically learned by the network just from the examples it is given.

\subsection{Regime dominated by baryonic effects}

Finally, we train neural networks with power spectra from the AstroCon set, setting the value of $k_{\rm pivot}$ to $0.5~h{\rm Mpc}^{-1}$, as above. We show the error achieved by that network on estimating $A$ and $B$ in the right panel of Fig.~\ref{fig:Toy_model} with blue lines. As expected, for $k_{\rm max}\leqslant k_{\rm pivot}$, the network is able to determine the parameters with the same accuracy as the networks trained using power spectra from the AstroNone and AstroDis sets.

For $k\geqslant k_{\rm pivot}$, we find that the network can determine the value of the cosmological parameters more accurately as we increase $k_{\rm max}$. This behavior is different from what we observed using power spectra from the AstroDis set. The reason is that in the regime dominated by baryonic effects, the power spectrum follows a law $P(k)=Ck^D$, but while the value of $D$ is not related to the value of the cosmological parameters, the value of $C=\bar{C}=Ak_{\rm pivot}^{B-D}$ is. The higher the value of $k_{\rm max}$, the better the network can constraint $C$ and $D$, and therefore the more information it can pull to determine $A$ and $B$. This shows how neural networks can extract cosmological information that is buried in the regime dominated by baryonic effects. 

We conclude this section by summarizing our findings with the power spectra. We find that neural networks can be trained to 1) find an optimal unbiased estimator that allows to extract the maximum cosmological information available, 2) marginalize the scales that are affected by baryonic effects, and 3) extract cosmological information that is buried in the regime dominated by baryonic effects. These conclusions are derived from the analysis on the data from our toy model I: power spectra. We now investigate whether these conclusions hold for a more complex problem: 2D density fields.

\section{Toy model II: Gaussian density fields}
\label{sec:Toy_model_II}

\begin{figure*}
\centering
\includegraphics[width=0.99\textwidth]{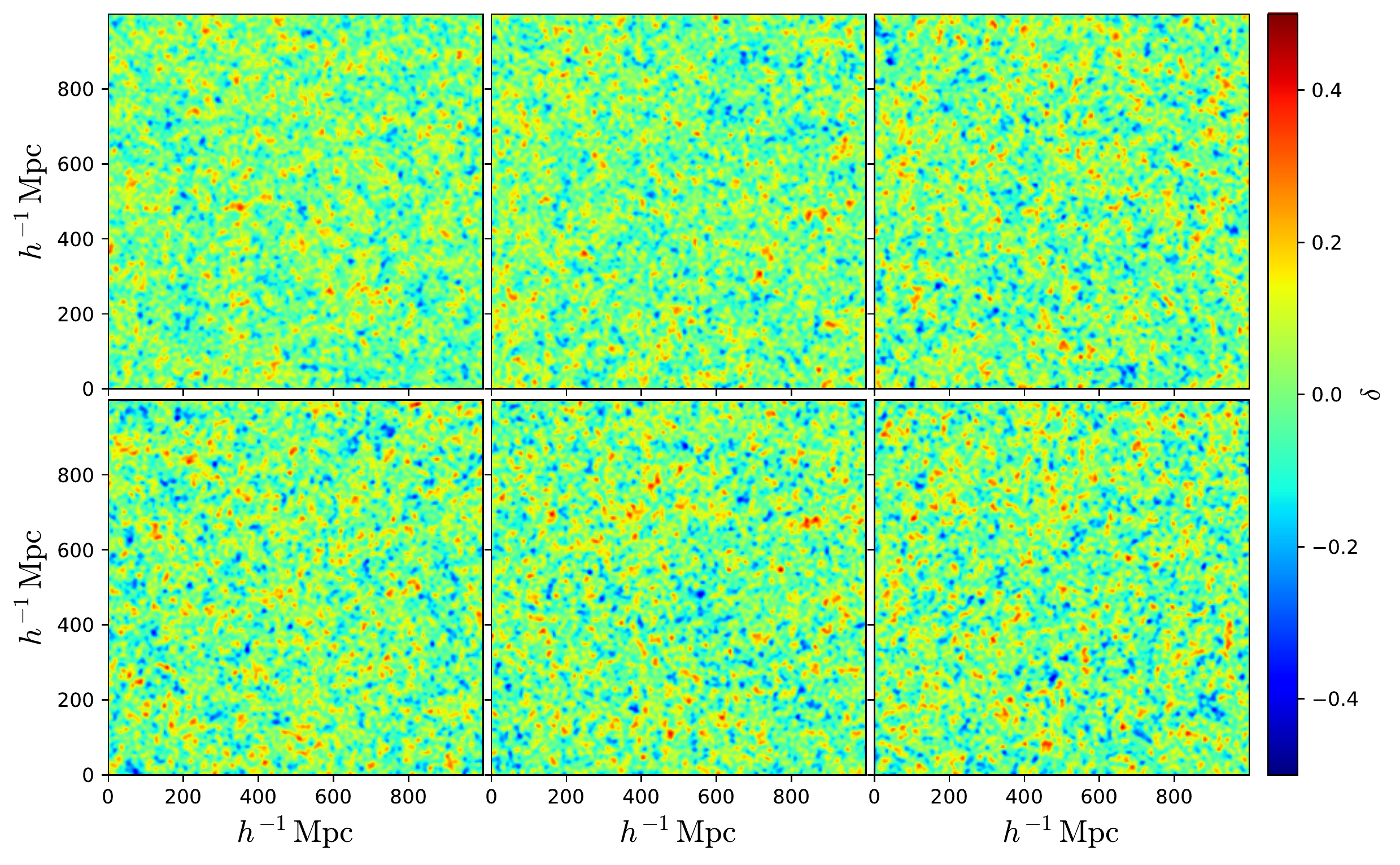}
\caption{We generate 2D Gaussian density fields with $128\times128$ pixels. The power spectrum of the maps is given by $P(k)=A/\sqrt{k}$, where $A$ is a free, \textit{cosmological}, parameter. Our goal is to train neural networks to predict the value of $A$ from these density fields. This plot shows 6 examples of such density fields. The value of $A$ is approximately 0.8, 0.9, 1.0, 1.0, 1.1 and 1.2, from top-left to bottom-right. All these maps belong to the AstroNone dataset, i.e. they are not affected by baryonic effects. We shall see that neural networks can determine the value of $A$ with an accuracy of $\simeq1\%$ from these maps.}
\label{fig:Gaussian_field_maps}
\end{figure*}

In this section, we repeat the analysis carried out for the toy model I, but using more complex and rich data: 2D Gaussian density fields. We made use of Gaussian density fields since their statistical properties can be fully characterized by their power spectrum. This is very useful, as it allows us to quantify the maximum information content of these fields in a simple and robust way. In other words, for these fields, we know the optimal estimator to extract the maximum information and we can write its likelihood. Our goal is to train neural networks to predict the value of cosmological parameters from 2D Gaussian density fields that may or may not be affected by baryonic effects.

\subsection{Data}
\label{subsec:dataII}

A generic 2D density field can be characterized by the value of its density contrast, $\delta(\vec{x})=\rho(\vec{x})/\bar{\rho}-1$, or by its Fourier transform $\delta(\vec{k})$
\begin{equation}
\delta(\vec{k})=\int d^2\vec{x}e^{-i\vec{k}\cdot\vec{x}}\delta(\vec{x})~.
\end{equation}
For each mode $\vec{k}$, $\delta(\vec{k})$ is a complex number, and therefore can be written as
\begin{equation}
\delta(\vec{k})=\alpha_{\vec{k}}e^{i\theta_{\vec{k}}}
\end{equation}
where $\alpha_{\vec{k}}$ and $\theta_{\vec{k}}$ are the mode's amplitude and phase, respectively. In a Gaussian density field, $\theta_{\vec{k}}$ follows a uniform distribution between 0 and $2\pi$, while $y=\alpha_{\vec{k}}$ follows a Rayleigh distribution
\begin{equation}
p(y)dy=\frac{y}{\sigma^2}e^{-y^2/2\sigma^2}dy
\label{eq:Rayleigh_distribution}
\end{equation}
where $\sigma^2=\beta P(k)/(8\pi^2)$, with $\beta$ is the area covered by the density field, and $P(k)$ the power spectrum. $p(y)dy$ denotes the probability that the value of $y$ is in the interval $[y,y+dy]$.

\begin{table*}
\begin{center}
\renewcommand{\arraystretch}{0.7}
\resizebox{0.95\textwidth}{!}{\begin{tabular}{| c || c | c | c | c | c | c | c| }
\hline
\multirow{3}{*}{Name} & Baryonic & \multirow{3}{*}{$P(k)$} & \multirow{3}{*}{Number of maps / Usage}  & \multirow{3}{*}{$A$} & \multirow{3}{*}{$D$} & \multirow{3}{*}{$C$} & $k_{\rm pivot}$ \\[0.5ex]
& effects? & & & & &&[$h{\rm Mpc}^{-1}$]\\[0.5ex]
\hline \hline
\multirow{5}{*}{AstroNone} & \multirow{5}{*}{No} & \multirow{5}{*}{$A/\sqrt{k}$} & 70,000 / Training & \multirow{5}{*}{[0.8 , 1.2]} & \multirow{5}{*}{-} & \multirow{5}{*}{-} & \multirow{5}{*}{-} \\[0.5ex] 
&&&  15,000 / Validation & & &  &\\[0.5ex]
&&&  15,000 / Testing & & & & \\[0.5ex]
\hline
AstroNone0.8 & No & $A/\sqrt{k}$ & 100,000 / Testing  & 0.8 & - & - & - \\[0.5ex]
\hline
AstroNone0.9 & No &$A/\sqrt{k}$ & 100,000 / Testing  & 0.9 & - & - & -  \\[0.5ex]
\hline
AstroNone1.0 & No & $A/\sqrt{k}$ & 100,000 / Testing  & 1.0 & - & - & - \\[0.5ex]
\hline
AstroNone1.1 & No & $A/\sqrt{k}$ & 100,000 / Testing  & 1.1 & - & - & -  \\[0.5ex]
\hline
AstroNone1.2 & No & $A/\sqrt{k}$ & 100,000 / Testing  & 1.2 & - & - & -  \\[0.5ex]
\hline
\multirow{5}{*}{AstroCon} & \multirow{5}{*}{Yes} & \multirow{2}{*}{$A/\sqrt{k}$ if $k\leq k_{\rm pivot}$} & 70,000 / Training & \multirow{5}{*}{[0.8 , 1.2]} & \multirow{5}{*}{[-1.0 , 1.0]} & \multirow{5}{*}{$\bar{C}$} & \multirow{5}{*}{0.3} \\[0.5ex] 
&&\multirow{3}{*}{$Ck^D$ if $k>k_{\rm pivot}$} &  15,000 /  Validation & &&&  \\[0.5ex]
&&&  15,000 / Testing &  &&&  \\[0.5ex]
\hline
\multirow{5}{*}{AstroDis} & \multirow{5}{*}{Yes} & \multirow{2}{*}{$A/\sqrt{k}$ if $k\leq k_{\rm pivot}$} &  70,000 / Training & \multirow{5}{*}{[0.8 , 1.2]} & \multirow{5}{*}{[-1.0 , 1.0]} & \multirow{5}{*}{[0.7 , 1.3]$\bar{C}$} & \multirow{5}{*}{0.3} \\[0.5ex] 
& &\multirow{3}{*}{$Ck^D$ if $k>k_{\rm pivot}$} &  15,000 / Validation &  & && \\[0.5ex]
&  &&  15,000 / Testing  &  & && \\[0.5ex]
\hline
\end{tabular}}
\end{center}
\caption{Summary of the different datasets used when working with the 2D Gaussian density fields (toy model II). $A$ represents the value of the \textit{cosmological parameter}, while $C$ and $D$ are the \textit{astrophysics parameters}, controlling the amplitude and shape of the power spectrum on scales $k>k_{\rm pivot}$, where baryonic effects show up. $\bar{C}$ is defined as the value of $C$ that makes the power spectrum continuous at $k_{\rm pivot}$: $A\sqrt{k_{\rm pivot}}=\bar{C}k_{\rm pivot}^D$. All maps from the different datasets have a different value of the initial random seed. Numbers in brackets indicate that the value of that parameter is randomly chosen from a uniform distribution within the quoted values.}
\label{table:data_Gaussian_fields}
\end{table*}

The way we construct the 2D Gaussian density fields is as follows. First, we need an input power spectrum, $P(k)$, that will characterize the Gaussian density field. We then populate the density field modes. For each mode, $\delta(\vec{k})=\alpha_{\vec{k}}e^{i\theta_{\vec{k}}}$, we select the value of $\theta_{\vec{k}}$ by taking a random number between 0 and $2\pi$, with uniform sampling. The mode amplitude, $\alpha_{\vec{k}}$, is drawn from the distribution of Eq. \ref{eq:Rayleigh_distribution}, which implicitly depends on the amplitude of the power spectrum at the wavenumber $k=|\vec{k}|$. When populating the modes in Fourier space it is very important to fulfill the Hermitian condition, $\delta(-\vec{k})=\delta(\vec{k})^*$, which arises from the fact that the Gaussian density field in configuration space is real, i.e. $\delta(\vec{x})^*=\delta(\vec{x})$. Finally, we make a Fourier transform to obtain the Gaussian density field in configuration space. The random numbers used to draw the mode's amplitude and phases can be recovered from an initial integer number, called the initial random seed. In this paper we work with Gaussian density fields containing $128\times128$ pixels and simulating an area of $1~(h^{-1}{\rm Gpc})^2$.

Similarly to toy model I, we consider three different sets of 2D Gaussian density fields:

\begin{itemize}
    \item \textbf{AstroNone}. These maps are not affected by baryonic effects. Thus, their underlying power spectrum is simply given by $P(k)=A/\sqrt{k}$. 
    \item \textbf{AstroDis}. Maps in this dataset are affected by baryonic effects on scales $k>k_{\rm pivot}$. As for the case of the toy model I, we model baryonic effects as a change in the amplitude and shape of the power spectrum on those scales. In these maps, the underlying power spectrum is given by $P(k)=A/\sqrt{k}$ for scales $k\leq k_{\rm pivot}$, and as $P(k)=Ck^D$ for $k>k_{\rm pivot}$. The power spectrum is not required to be continuous at $k_{\rm pivot}$.
    \item \textbf{AstroCon}. Maps in this dataset are affected by baryonic effects that are modelled in the same way as for AstroDis, with the only difference being that these maps are required to have a continuous power spectrum, i.e. $A/\sqrt{k_{\rm pivot}}=Ck_{\rm pivot}^D$.
\end{itemize}

We note that for simplicity and to keep the data as small and interpretable as possible, we have considered a single \textit{cosmological} parameter $A$. We showed in the previous section that neural networks can find the optimal solution also in presence of several, correlated, variables. Our conclusions thus do not depend on this choice.

Although the generation of these maps is very computationally efficient, it is not fast enough to generate them on the fly while training the neural networks. Thus, we create different catalogues containing Gaussian density fields from the different datasets. We create 100,000 AstroNone, 100,000 AstroDis, and 100,000 AstroCon maps. Within each of those sets, we split the maps into 70,000, 15,000, and 15,000 subsets that we use for training, validation, and testing, respectively. In all cases, the value of $A$ is taken by sampling a uniform distribution from $0.8$ to 1.2. For AstroCon and AstroDis, the value of $D$ is taken from a uniform distribution between -1 and +1. In the case of AstroCon, $C$ is fixed to $\bar{C}=A/k_{\rm pivot}^{0.5+D}$, while for AstroDis its value is taken by randomly sampling a uniform distribution between $0.7\bar{C}$ and $1.3\bar{C}$.

We have also generated a set of catalogues containing maps with a fixed value of $A$: 0.8, 0.9, 1.0, 1.1, and 1.2. Each of these catalogues contains 100,000 maps and we use them to test the network and to compare the variance of the estimator learned by the network against the variance of the optimal estimator from the Fisher matrix. We emphasize that all the Gaussian maps generated have a different value of the initial random seed. 

A summary with the different sets and their characteristics is shown in Table \ref{table:data_Gaussian_fields}. We show six examples of the generated Gaussian density fields in Fig.~\ref{fig:Gaussian_field_maps}.

\subsection{Neural networks}
\label{subsec:NN2}

We train a model that combines convolutional with fully connected layers; the details on the architecture we use are outlined in Appendix \ref{sec:Architectures}. We use the Adam optimizer with a learning rate of $10^{-5}$, with values of the beta parameters equal to $\vec{\beta}=\{ 0.5, 0.999\}$. We use as loss function the standard mean square error:
\begin{equation}
    \mathcal{L}=\frac{1}{N}\sum_{i=1}^N(A_{\rm NN}-A_{\rm true})^2
\end{equation}
where $A_{\rm true}$ and $A_{\rm NN}$ are the true and predicted values of the cosmological parameter $A$. The sum runs over all maps in the training set. We do not make use of dropout, but we use a value of the weight decay equal to $2\times10^{-4}$. We train the network using a batch size equal to 128. We did not performed data augmentation while training the network, as we find that our dataset is large enough to avoid concerns with overfitting. We train the network for approximately 6,000 epochs. When the loss plateaus, we decrease the learning rate by a factor of 10, and continue training.

\begin{figure}
\centering
\includegraphics[width=0.49\textwidth]{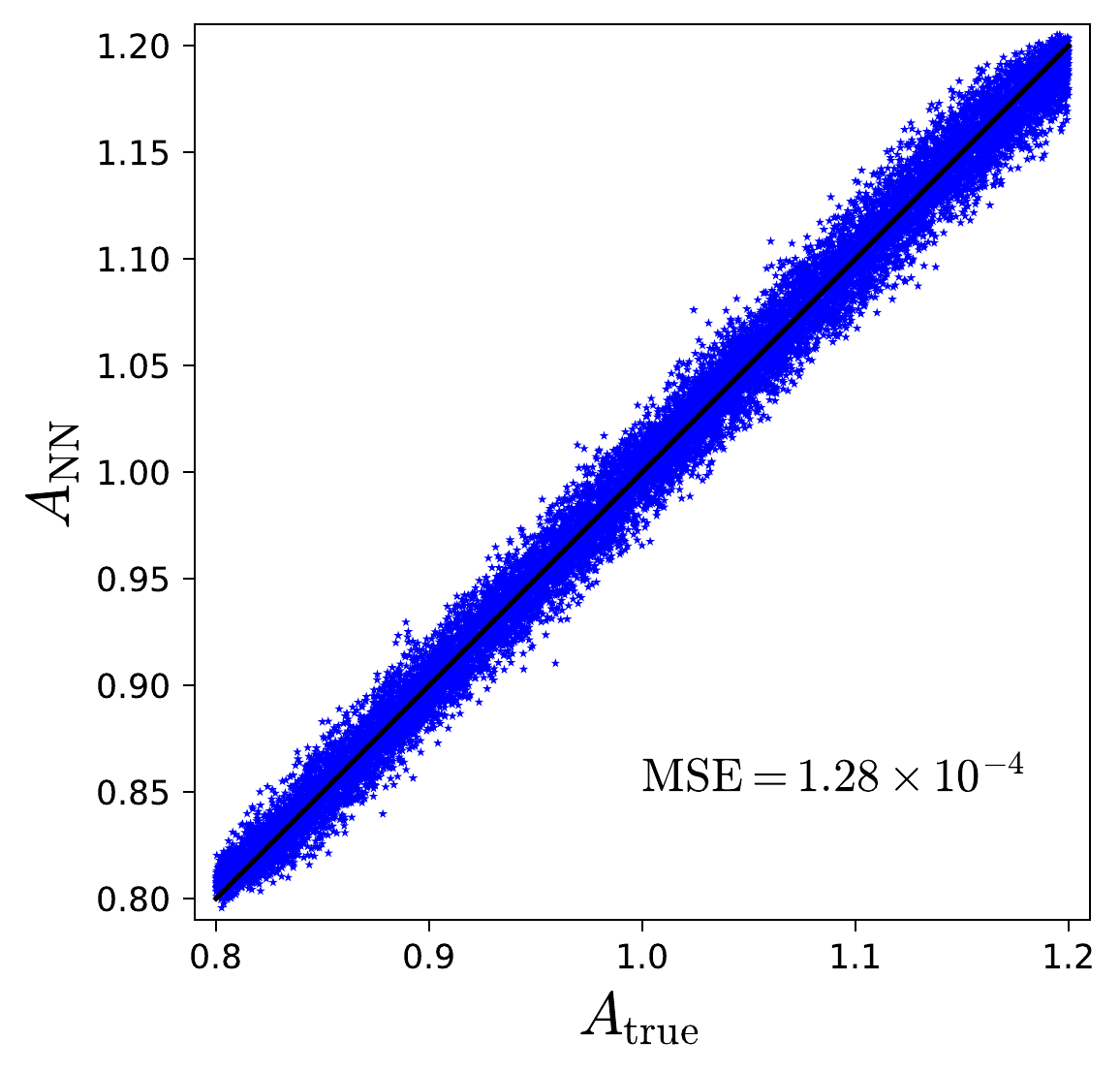}
\caption{We train a convolutional neural network to predict the value of the cosmological parameter $A$ from 2D Gaussian density fields. This plot shows the predicted value of $A$ as a function of its true value for the 15,000 Gaussian maps in the AstroNone test set. The solid black line indicates a perfect agreement, i.e.~$A_{\rm NN}=A_{\rm true}$. The network is able to predict the value of $A$ with a high accuracy (RMSE) of 0.0113, while the expected maximum performance from the Fisher matrix will be 0.0111.}
\label{fig:A_NN_vs_A_true}
\end{figure}

\begin{figure*}
\centering
\includegraphics[width=0.99\textwidth]{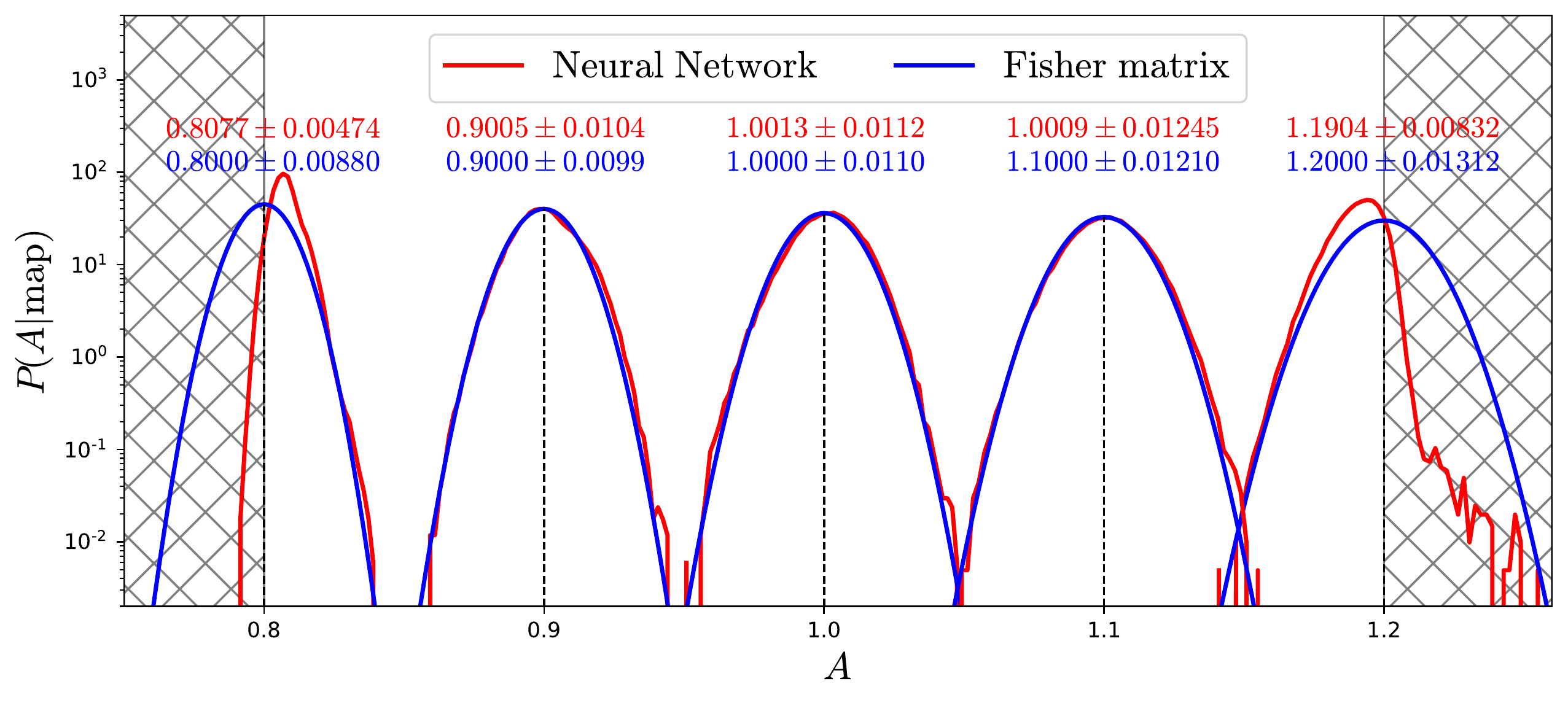}
\caption{We train a neural network to predict the value of the cosmological parameter $A$ from 2D Gaussian density fields of the AstroNone dataset. Once the network has been trained, we use it to predict the value of the cosmological parameter from  the AstroNone0.8, AstroNone0.9, AstroNone1.0, AstroNone1.1, and AstroNone1.2 datasets. The red lines show the distribution function of the values of $A$ predicted by the network, while the blue line represents the optimal bounds from the Fisher matrix calculation. The numbers on the top show the mean and standard deviation.  As can be seen, the agreement is excellent; the network is able to get an unbiased value of $A$ with an error that is only $\mathcal{O}(1\%)$ worst than the one from the optimal estimator. The hatch areas represent regions of the parameter space not shown to the network when training it. Near those regions, the network behaves as a biased estimator. This happens because the network learns the priors of the distribution it has been trained on.}
\label{fig:Posterior_Gaussian_fields}
\end{figure*}

Once the network is trained, we feed it with the 15,000 Gaussian maps of the AstroNone test set. Fig.~\ref{fig:A_NN_vs_A_true} shows the predicted values of $A$ as a function of their true values. We can see that the network is able to predict the value of the cosmological parameter with a high accuracy: the MSE and RMSE are $1.28\times10^{-4}$ and $0.0113$, respectively. Furthermore, from visual inspection it seems that the network has found an unbiased estimator (besides for low or high values of $A$). Next we compare these results against the variance of the optimal estimator that we obtain by employing the Fisher matrix formalism.

\subsection{Optimal estimator}

The statistical properties of Gaussian density fields can be fully characterized by their power spectrum. In other words, the power spectrum is the optimal estimator to extract information from Gaussian density fields. Thus, if we want to quantify how accurately we can constrain the value of some parameters from Gaussian density fields, we can simplify the question as: how well can the power spectrum determine the value of these parameters?

\begin{figure*}
\centering
\includegraphics[width=0.99\textwidth]{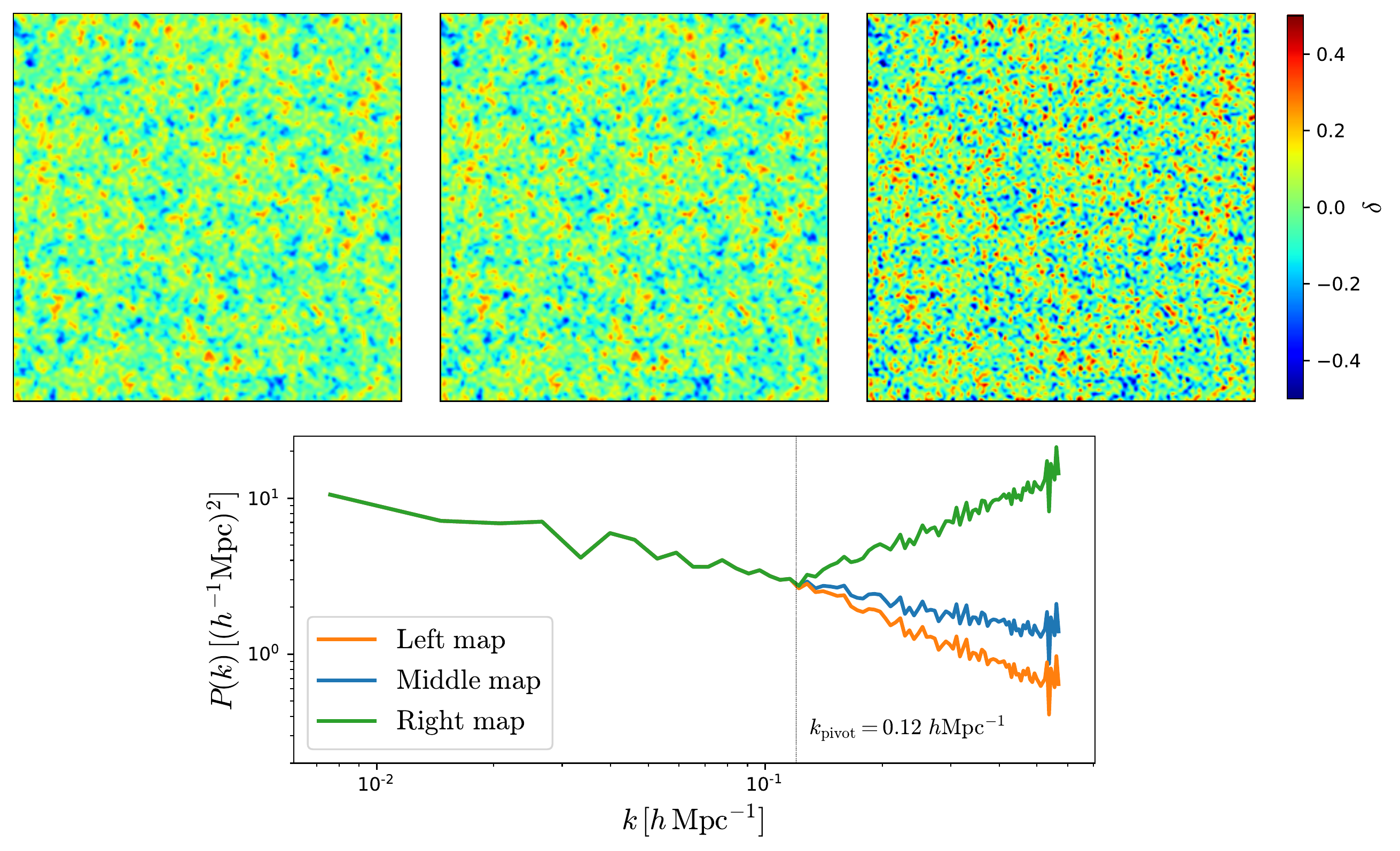}
\caption{The upper panels show three Gaussian density fields. The maps have the same mode phases, but differ in their mode amplitudes. The bottom panel displays their power spectra. While the three fields have the same power on large scales, they differ on small scales: the map on the right has the highest power on small scales, while the left map has the lowest.}
\label{fig:Maps_astrophysics}
\end{figure*}

In this case, we use a different formalism to the one employed for toy model I, and instead of finding the value of $A$ that maximizes the likelihood, we use the Fisher matrix formalism to quantify the variance of the optimal estimator. While using the maximum likelihood method employed in the previous section will not change our conclusions, we use this alternative method to probe that the estimator learned by the network is the one with the lowest variance.

We now briefly outline the Fisher matrix formalism. Consider a given statistic $\vec{S}=\{S_0, S_1,...,S_N\}$, and some parameters $\vec{\theta}=\{\theta_0, \theta_1,...,\theta_M\}$. the Cramer-Rao bound \citep{Cramer, Rao} states that the error on the parameter $\theta_i$, $\sigma_{\theta_i}$ from an optimal unbiased estimator saturates at
\begin{equation}
    \sigma_{\theta_i} \geq \sqrt{F_{ii}^{-1}}~,
\end{equation}
where $F_{ij}$ is the Fisher matrix
\begin{equation}
    F_{ij}=\frac{\partial S_\alpha}{\partial \theta_i}C^{-1}_{\alpha \beta}\frac{\partial S_\beta}{\partial \theta_j}~,
\end{equation}
and $C^{-1}_{\alpha \beta}$ is the covariance matrix. In our case, the statistic used is the power spectrum, $P(k)=A/\sqrt{k}$, while the only parameter considered is $A$. Both the derivatives and the covariance are trivial
\begin{eqnarray}
\frac{\partial P(k_\alpha)}{\partial A} &=& \frac{P(k_\alpha)}{A}\\
C_{\alpha \beta}&=&\frac{2P(k_\alpha)^2}{N_{k_\alpha}}\delta_{\alpha \beta}
\end{eqnarray}
where $N_{k_\alpha}$ is the number of modes in the $k_\alpha$ bin, and $\delta_{\alpha \beta}$ is the Kronecker delta. The Fisher matrix (which in this case is a scalar) thus reduces to
\begin{equation}
    F=\frac{1}{2A^2}\sum_{\alpha} N_{k_\alpha}~.
    \label{Eq:Fisher_error2}
\end{equation}
We note that the above sum goes through all $k$-bins in the power spectrum. That sum should thus be equal to the total number of pixels in the Gaussian field, $N_{\rm pixels}$. The error on the parameter $A$ is finally given by
\begin{equation}
    \sigma_A\geq A\sqrt{\frac{2}{N_{\rm pixels}}}~.
    \label{Eq:Fisher_error}
\end{equation}
The above expression provides the lower bound on the square root of the variance of the optimal unbiased estimator when the fiducial value of the cosmological parameter is $A$. We may be interested in the average error on the parameter $A$ when $A$ varies within a given range $A\in[A_{\rm min} - A_{\rm max}]$. That error can be computed as
\begin{equation}
    \bar{\sigma}_A=\sqrt{\frac{\int_{A_{\rm min}}^{A_{\rm max}}\sigma_A^2~dA}{\int_{A_{\rm min}}^{A_{\rm max}} dA}}=\sqrt{\frac{A_{\rm max}^2+A_{\rm max}A_{\rm min}+A_{\rm min}^2}{1.5N_{\rm pixels}}}~.
    \label{Eq:Fisher_error_mean}
\end{equation}
For $A_{\rm max}=1.2$ and $A_{\rm min}=0.8$, and for Gaussian fields with $128\times128$ pixels, the above expression yields: $\bar{\sigma}_A=0.0111$. 

That number can be directly compared with the RMSE achieved by the neural network over the same A range: 0.0113. Our neural network behaves as an estimator of the parameter $A$ whose variance is only $1.8\%$ larger than the one of the optimal estimator. It is feasible to further decrease the error on the neural network and get closer to the Fisher results, e.g.~with more hyperparameter tuning, an improved model architecture, or more training data. We emphasize that the network did not know, or was informed, that the data it was trained on, were Gaussian density fields.

We now investigate the effect of priors by comparing the prediction of the network against the expectation from the Fisher analysis for maps with fixed value of the cosmological parameter $A$. We made use of the AstroNone1.0 dataset, where all maps have a value of $A$ equal to 1. We feed each map into the neural network, and obtain the value of $A$ predicted by the network. We then compute the distribution of the values of $A$. 

The Fisher expectation can be obtained as follows. First, the error on the parameter for its fiducial value, $A_{\rm fid}$, can be calculated using Eq. \ref{Eq:Fisher_error}, and the distribution of the parameter $A$ is expected to follow a Gaussian distribution with mean $A_{\rm fid}$ and standard deviation $\sigma_A$
\begin{equation}
    p(A)dA = \frac{1}{\sqrt{2\pi\sigma_A^2}}\exp\left(-\frac{(A-A_{\rm fid})^2}{2\sigma_A^2}\right)dA~.
\end{equation}
We show the results of this analysis in Fig.~\ref{fig:Posterior_Gaussian_fields}; blue lines show Fisher bounds while red lines are the distributions from the neural network predictions. 

When the value of $A_{\rm fid}$ is equal to 1, the agreement between the Fisher and the neural network is very good. The means and standard deviations of the two distributions are: $1.0013\pm0.0112$ (network) and $1.0000\pm0.0110$ (Fisher). We find that the neural network behaves as an optimal unbiased estimator of the parameter $A$: its standard deviation is only $\sim1.8\%$ higher than the one from the theoretical optimal estimator. 

We have repeated the above exercise for maps with different values of $A_{\rm fid}$: 0.8, 0.9, 1.1, and 1.2, using the AstroNone0.8, AstroNone0.9, AstroNone1.1, and AstroNone1.2, respectively. We show the results in Fig.~\ref{fig:Posterior_Gaussian_fields}. For values of $A_{\rm fid}$ equal to 0.9 and 1.1, we reach similar conclusions as for $A_{\rm fid}=1.0$: the network has found an unbiased estimator that achieves almost the same error as the optimal estimator: $5\%$ and $3\%$ larger errors than Fisher for $A_{\rm fid}$ equal to 0.9 and 1.1, respectively. We emphasize that the error on the parameter $A$ depends on its fiducial value (see Eq. \ref{Eq:Fisher_error}). This dependence is automatically incorporated into the neural network.

For values of $A_{\rm ref}$ equal to 0.8 and 1.2, the network provides a distribution of values of $A$ that significantly differs from the optimal one from Fisher. Not only the mean value is biased, but the width of the distribution is smaller than the one expected. We note that the network biases the value of $A$ towards small/high values when $A_{\rm fid}$ is high/small. This happens because the network has, at least partially, learned the priors. In other words, the network has never seen a Gaussian map whose value of $A$ is larger/smaller than 1.2/0.8, and is making use of that information. As can be seen, in this regime, the network behaves as a biased estimator of $A$ that has a lower variance than the optimal one from Fisher. As we found for the cases of the Toy model I, the effects of the priors can produce that the network finds an estimator with a lower variance than the theoretical floor\footnote{We emphasize that this happens because our theory calculation did not include the priors. If we would have included the priors on the Fisher matrix calculation, its variance would have been lower than the one of the network.}; this happens at the expenses of being a biased estimator.

We thus conclude that for values of $A$ sufficiently far away of the training boundaries, the neural network has learned an unbiased estimator to determine the value of $A$. It automatically includes the dependence of $\sigma_A$ with $A$. The comparison with the error from the optimal unbiased estimator from the Fisher matrix calculation shows that the learned estimator is almost optimal. However, near the edges of the interval where the network has been trained, we observe some effects that indicate that the network may be make use of additional information from priors.

\subsection{Baryonic effects}
\label{subsec:astrophysics_maps}

We now investigate the effect of contaminating Gaussian density fields with baryonic effects. To this end, we train a neural network to predict the value of the cosmological parameter $A$ when the input are Gaussian density fields that are generated from power spectra with the shape
\[ P(k)=\begin{cases} 
      A/\sqrt{k} & {\rm if}~ k\leq k_{\rm pivot} \\[1ex]
      Ck^D & {\rm if}~ k>k_{\rm pivot}~.
   \end{cases}
  \label{Eq:Pk_astrophysics}
\]

We illustrate this in Fig.~\ref{fig:Maps_astrophysics}, where we show three Gaussian density fields that have the same value of $A$ and the initial random seed, but different values of $D$. For this example, we have set $k_{\rm pivot}=0.12~h{\rm Mpc}^{-1}$ and we have considered three different values of $D$: -1, -0.5, and 1. In these maps $C$ is determined by requiring that the power spectrum is a continuous function (AstroCon set): $A\sqrt{k_{\rm pivot}}=Ck_{\rm pivot}^D$. It can easily be appreciated that the amplitude of the power spectrum on small scales can produce large visual differences on the maps.

 We train a neural network to predict the value of $A$ from maps of the AstroDis set, i.e. Gaussian maps contaminated by baryonic effects on scales $k>k_{\rm pivot}$. The architecture, setup, and training procedure are identical to the neural network presented in Section \ref{subsec:NN2}, with the only difference being the value of the weight decay, which is set to $4\times10^{-4}$.

\begin{figure*}
\centering
\includegraphics[width=0.79\textwidth]{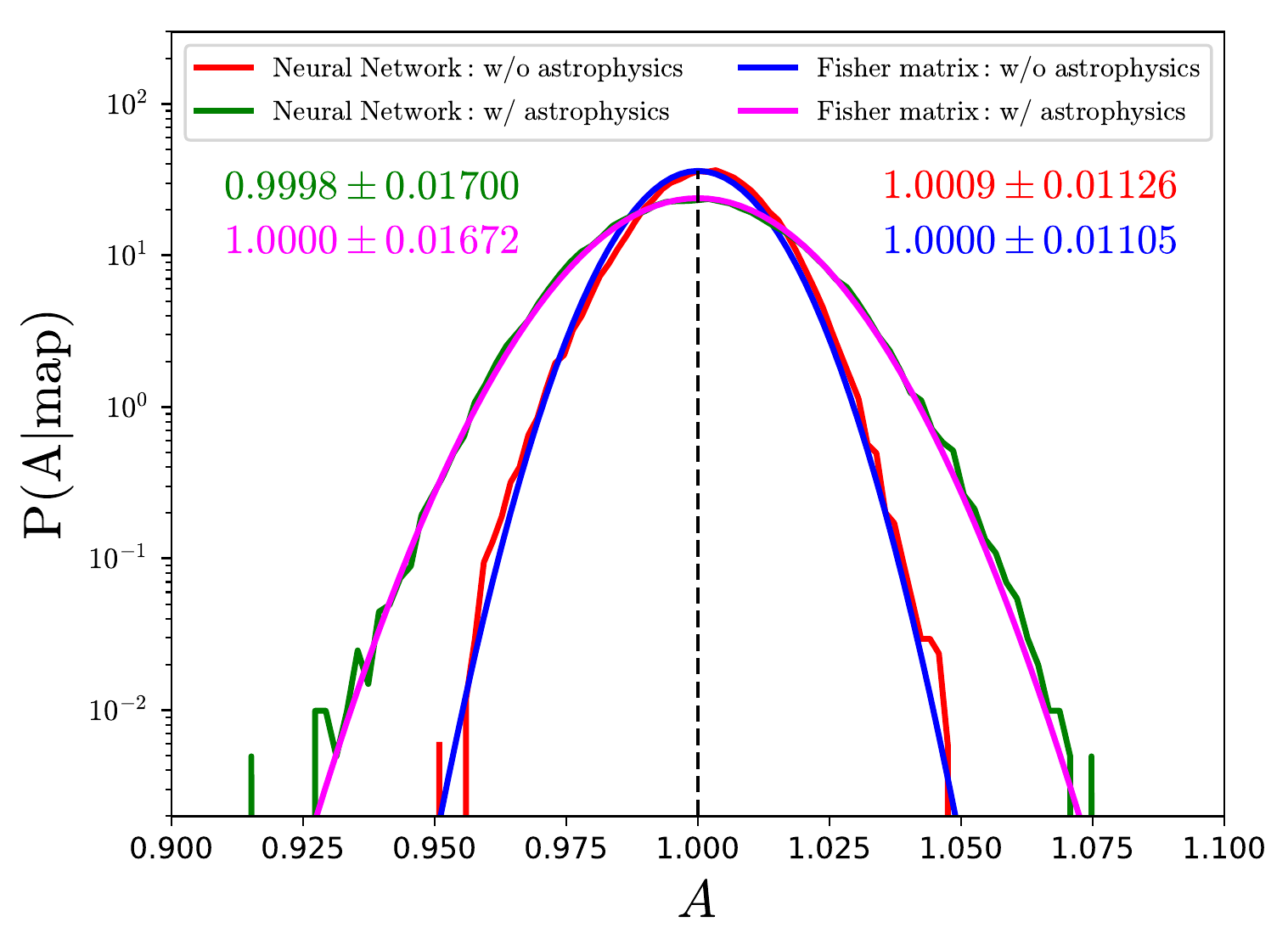}
\caption{We have trained neural networks to predict the value of the cosmological parameter $A$ from 2D Gaussian density fields. One network is trained using maps with no baryonic effects, while the other network is fed with maps contaminated with baryonic effects at $k > k_{\rm pivot}=0.3~h{\rm Mpc}^{-1}$. Once the networks have been trained, we input to them 100,000 Gaussian maps of the AstroNone1.0 set that have a true value of $A=1$. The red and green lines show the distribution of the values predicted by the network. As expected, the width of the distribution for the maps affected by baryons is higher, showing that baryonic effects erase some information. The blue and magenta lines show the distribution on $A$ derived from a Fisher matrix calculation assuming that the cosmological information is located on all scales and on scales larger than $k_{\rm pivot}$, respectively. The agreement between the distributions from the networks and optimal unbiased estimator from the Fisher matrix is excellent, pointing out that the neural network has learned an optimal estimator that extracts all cosmological information while marginalizing over baryonic effects. The numbers in the interior of the plot indicate the mean and standard deviation of the different distributions.}
\label{fig:Posterior_astrophysics}
\end{figure*}

Once the network is trained, we use the 15,000 maps of the AstroDis test set to determine its accuracy. The network achieves a MSE=$2.838\times10^{-4}$ and RMSE=$0.0168$. We note that these numbers are worse than those obtained from the maps that did not incorporate baryonic effects, as expected. 

We now compare the network performance against the error from the optimal estimator using the Fisher matrix. For this, we need to take into account that we have created the 2D Gaussian maps of the AstroDis dataset in such a way that the power spectrum at $k>k_{\rm pivot}$ has no information about the clustering pattern on larger scales\footnote{Strictly speaking, the value of $C$ is drawn from the value of $\bar{C}$, that knows about cosmology. However, the priors are so large that in practice the value of $C$ can be considered as independent of the large-scale clustering.}. Thus, we expect that all cosmological information will reside in the regime where $k\leq k_{\rm pivot}$. We can thus quantify the maximum cosmological information embedded into these fields by using the Fisher formalism. In this case we made use of Eq. \ref{Eq:Fisher_error2} and cut the sum for modes with $k<k_{\rm pivot}$. For maps with $128\times128$ pixels and a side length of $1000~h^{-1}{\rm Mpc}$ there are 7,154 modes with $k\leq0.3~h{\rm Mpc}^{-1}$, yielding an error estimate on $A$ for an optimal unbiased estimator of
\begin{equation}
    \sigma_A=0.0167A~.
\end{equation}
If we consider the mean error in the range $A\in[0.8 - 1.2]$, following the same procedure used to derive Eq.~\ref{Eq:Fisher_error_mean}, we obtain
\begin{equation}
    \bar{\sigma}_A=0.0169~.
\end{equation}
This number can be directly compared with the RMSE from the network: 0.0168. Thus, the agreement between the prediction of the neural network and the Fisher, in terms of constraining power on the parameter $A$, is excellent. We note that in this case the network is slightly outperforming the Fisher matrix. This happens because the network accounts for priors effects near its boundaries, while we did not take this into account in the Fisher matrix calculation. We have repeated the above exercise for different values of $k_{\rm pivot}$, reaching similar conclusions\footnote{For small values of $k_{\rm pivot}$ we find that the network outperforms the Fisher matrix. This happens because the prior information becomes more important on those scales, as constraints on the parameters are quickly decreasing with decreasing $k_{\rm pivot}$, in the same fashion as with the Toy model I.}.

We now investigate the performance of the network in a bit more detail. Once the network is trained, we feed it with the 100,000 Gaussian density fields of the AstroNone1.0 dataset. We feed the network with these maps, that do not contain baryonic effects, to investigate if the network has learned to marginalize over scales smaller than $k_{\rm pivot}$. In Fig.~\ref{fig:Posterior_astrophysics} we show with a solid green line the distribution of the predicted value of $A$ from the neural network. The magenta line in that plot shows the Fisher expectation. As can be seen, the agreement is excellent. The mean and standard deviation from the two distributions are: $0.9998\pm0.01700$ (network) and $1.0000\pm0.01672$ (Fisher). The agreement in the error of $A$ is around $1.5\%$. This points out that the network has learned an almost optimal unbiased estimator; that estimator is one that has learned to marginalize over baryonic effects. This can be better visualized when comparing the results against those obtained when no baryonic effects are present, shown as blue and red curves in Fig.~\ref{fig:Posterior_astrophysics}. We have repeated the same exercise using maps contaminated by baryonic effects at $k_{\rm pivot}=0.5~h{\rm Mpc}^{-1}$, reaching the same conclusions as with the maps from the AstroNone1.0 dataset. 

\begin{figure*}
\centering\includegraphics[width=0.99\textwidth]{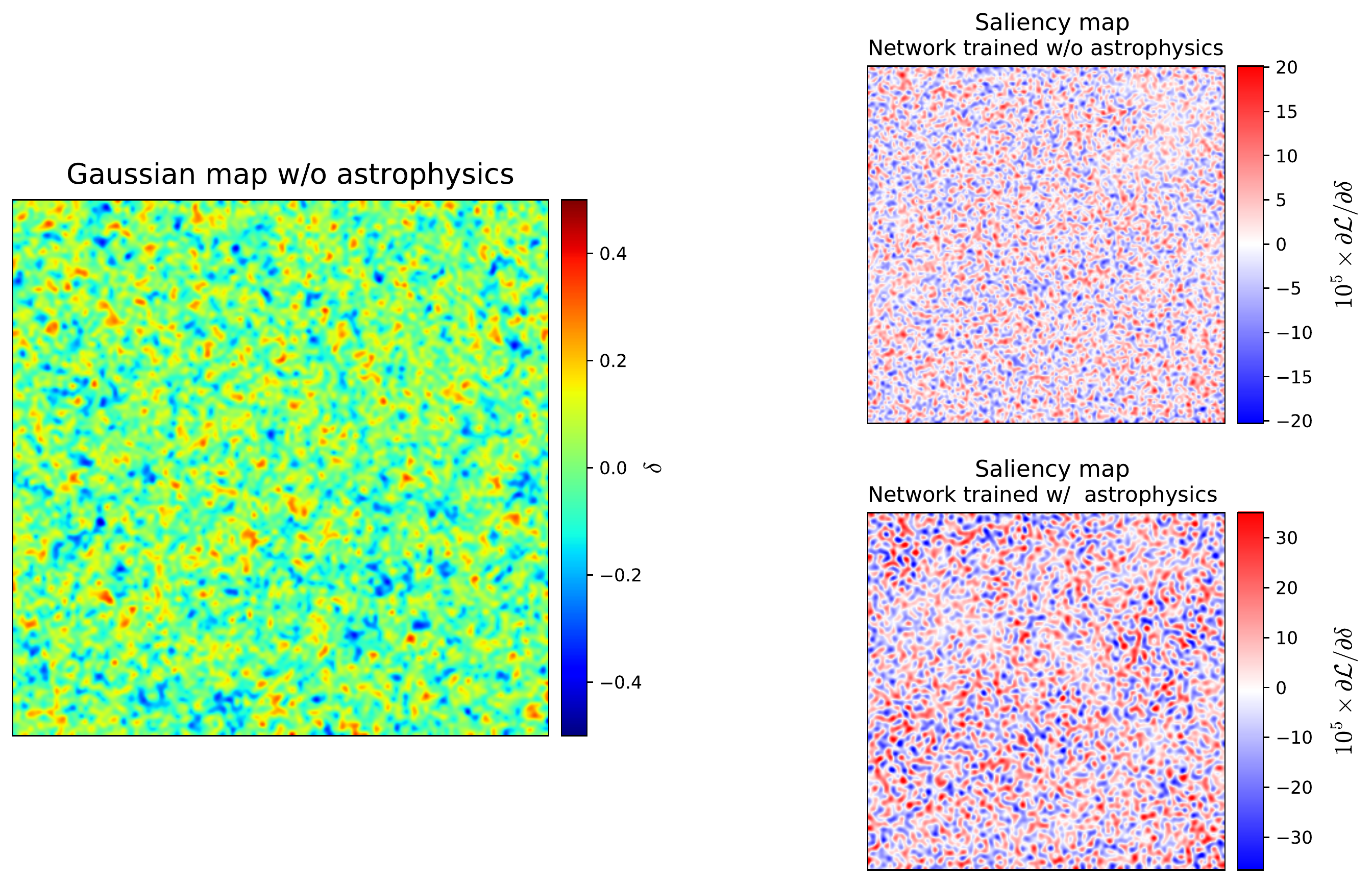}
\caption{We take the neural networks trained on maps with and without baryonic effects and feed them with the Gaussian map shown in the left panel. We then compute the saliency maps for the two networks, and show the results on the right panels. It can be seen that the saliency map of the network trained with maps contaminated by baryonic physics, has a lower resolution, i.e. there is no gradient information on small scales. This points out that the network trained with maps affected by baryonic effects is not looking at small scales when predicting the value of $A$, as expected.}
\label{fig:Saliency_astrophysics}
\end{figure*}

While the above results seem to clearly indicate that the network has learned to marginalize over the scales affected by baryonic effects, we would like to have a more direct probe of it. To show this more explicitly, we make use of saliency maps. Saliency maps are just the derivative of the loss function with respect to the input field. They can help with the interpretation of what the neural network is doing, since largest derivatives (in modulus) indicate more sensitive pixels/regions to the particular parameter considered.

We have taken a 2D Gaussian density field from the AstroNone test set and fed it to the two networks considered above; i.e.~networks trained using maps without and with baryonic effects. Next, we have computed the saliency map of each network. We show the results in Fig.~\ref{fig:Saliency_astrophysics}. The saliency map obtained from the network trained on maps affected by baryonic effects has a lower resolution: the observed features are coarser than those from the other network. This points out, in a more direct way, that the network has indeed learned to marginalize over the scales that are affected by baryonic effects, as na{\"i}vely expected.

\subsection{Regime dominated by baryonic effects}
\label{subsec:astrophysics_dominated_regime}

We now investigate the implications of using a continuous power spectrum when generating the Gaussian fields contaminated by baryonic effects. We train a neural network using maps from the AstroCon training set, whose underlying power spectra are required to be continuous. 

\begin{table*}
\begin{center}
\renewcommand{\arraystretch}{0.6}
\resizebox{0.69\textwidth}{!}{\begin{tabular}{| c || c | c | c |}
\hline
& \multicolumn{3}{|c|}{Neural Network trained on maps with} \\[0.5ex]
\hline
\multirow{2}{*}{Dataset} & \multirow{2}{*}{No baryonic effects} & Baryonic effects & Baryonic effects \\
& & $P(k)$ discontinuous & $P(k)$ continuous\\
\hline
AstroNone &  \multirow{2}{*}{$1.281\times10^{-4}$} &  \multirow{2}{*}{$2.838\times10^{-4}$} &  \multirow{2}{*}{$2.196\times10^{-4}$} \\
(test set) & & &\\
\hline
 \multirow{2}{*}{AstroNone1.0} &  \multirow{2}{*}{$1.269\times10^{-4}$} &  \multirow{2}{*}{$2.986\times10^{-4}$} & \multirow{2}{*}{$2.243\times10^{-4}$}\\
 & & & \\
\hline
\end{tabular}}
\end{center}
\caption{We have trained three neural networks using three different data sets: 1) maps with no baryonic effects (AstroNone), 2) maps with baryonic effects where the underlying power spectrum is discontinuous at $k_{\rm pivot}$ (AstroDis), and 3) maps with baryonic effects where the underlying power spectrum is continuous (AstroCon). The numbers in the table show the MSE predicted by the three networks when using maps with no baryonic effects with $A\in[0.8-1.2]$ (AstroNone), and for maps with no baryonic effects with a value of $A$ fixed to 1.0 (AstroNone1.0). We find that a network trained on maps with baryonic  effects and a continuous power spectrum can provide tighter constraints on the cosmological parameter than a network trained using maps with discontinuous baryonic  effects. This shows that the network has learned to extract cosmological information from scales $k>k_{\rm pivot}$, i.e. the regime completely dominated by baryonic effects.}
\label{table:results_astrophysics}
\end{table*}

The architecture of the network used is the same as in the previous cases. Once the network is trained, we evaluate its performance using the 15,000 maps of the AstroCon test set, achieving a mean squared error equal to $2.210\times10^{-4}$. We have also evaluated the performance of the network using the 100,000 maps of the AstroNone1.0 set; the MSE is $2.243\times10^{-4}$. We show the results of this analysis, and its comparison with the other networks, in Table \ref{table:results_astrophysics}. 

We find that the network trained on maps affected by baryonic effects and with a continuous power spectrum yield tighter constraints on $A$ than the network trained with maps with a discontinuous power spectrum (AstroDis). Given the tests performed in the previous subsection, this clearly indicates that the extra cosmological information the network is extracting arises from scales $k>k_{\rm pivot}$. In other words, the neural network has learned to extract the cosmological information embedded in the regime dominated by baryonic effects. 

We believe that what the network is doing is the following. Since the optimal estimator requires computing the power spectrum (or some equivalent quantity), the network may be computing that statistic from the maps. While the power spectrum on scales $k>k_{\rm pivot}$ is dominated by a power law, $Ck^D$, whose amplitude and shape are independent of the cosmological parameter $A$, there is a relation between these three parameters that is required for having a continuous power spectrum:
\begin{equation}
    A\sqrt{k_{\rm pivot}}=Ck_{\rm pivot}^D~.
\end{equation}
By going into the regime dominated by baryonic effects, the network can learn the values\footnote{We note that the deeper we go into this regime, the better we can constrain these parameters.} of $C$ and $D$. As we have seen in the previous subsection, the network can also learn the value of $k_{\rm pivot}$, so it can use the previous equation to better constrain the value of $A$. We emphasize that constraining the value of $A$ using the previous equation is a method completely different from determining its value from the clustering of the Gaussian density field. 

Using this larger, more complex and richer data set of 2D Gaussian density fields, we reach the same conclusion as with the power spectra of the toy model I. Neural networks can extract cosmological information that is buried in the regime dominated by baryonic effects, a regime that the network also learns to marginalize over.

\section{Summary}
\label{sec:Conclusions}


The most important findings of this paper can be summarized as follows:
\begin{itemize}
    \item Neural networks can find an optimal unbiased estimator that allows to extract the maximum information embedded into cosmological data.
    \item Neural networks can learn to marginalize over scales affected baryonic effects.
    \item Neural networks can extract cosmological information that may be buried in the regime dominated by baryonic effects.
\end{itemize}
We have reached the above conclusions by training neural networks with two different toy model datasets: 1) a summary statistic, the power spectrum, and 2) 2D Gaussian density fields. The reason behind using these simple datasets is that the optimal solution is known, which allows us to compare it against the results from the neural network.

In both cases, we have shown that neural networks learn the priors on the distribution they have been trained on; by construction, they are trained to find an approximation to the posterior mean. This may be a potential problem if the priors are comparable, or tighter, than the bounds on the parameters. A simple fix for this is to train the network over very broad parameter ranges. 

We emphasize that we have not provided the network with information about the structure of the data, e.g. whether the 2D maps are Gaussian density fields. The networks learned that by themselves. This is the reason why we believe that similar conclusions should be reached in the case of non-Gaussian density fields. Furthermore, in the case of data contaminated by baryonic effects, we never give information to the network on the scale where baryonic effects show up. The networks were able to learn that information just from the examples we fed them.

Our implementation of baryonic effects has been carried out using simplistic models. We however expect that our findings will hold for more complex, and realistic, implementations of the baryonic effects (e.g. from full numerical simulations). Furthermore, it has been shown that baryonic effects leave distinct signatures on different statistics; \citep[see][for the case of power spectrum and bispectrum]{Foreman_2019}. This opens the door to combining different statistics in a clever way that allows the extraction of cosmological information on the regime dominated by baryonic effects.

This paper justifies the approach followed recently by the CAMELS project \citep{CAMELS}. CAMELS is a suite of more than 4,000 state-of-the-art numerical simulations, run with thousands of different cosmological and astrophysical models using the baryonic subgrid physics implementations of the IllustrisTNG \citep{WeinbergerR_16a,PillepichA_16a} and SIMBA \citep{SIMBA} simulations, where several key parameters are varied across a wide range. One of the main goals of CAMELS is to train neural networks to extract the maximum cosmological information from 3D fields while marginalizing over astrophysical effects. 

\section*{ACKNOWLEDGEMENTS}
We thank Gabriella Contardo, Yin Li, Leander Thiele, Core Francisco Park, and Oliver Philcox for useful conversations. FVN acknowledge funding from the WFIRST program through NNG26PJ30C and NNN12AA01C. The work of BW, DAA, SG, SH, and DS has been supported by the Simons Foundation. DAA was supported in part by NSF grant AST-2009687. The code and analysis tools developed for this work are publicly available at \url{https://github.com/franciscovillaescusa/baryons_marginalization}. This work has made use of the Pylian3 libraries, publicly available at \url{https://github.com/franciscovillaescusa/Pylians3}.

\begin{appendix}

\section{A. Relation between the neural network estimator and the posterior mean}
\label{sec:posterior_mean}

In this appendix we show that the way we train neural networks guarantee that the estimator found approaches the posterior mean. Consider solving the following least square optimization problem
\begin{eqnarray}
I[f]&=&\int (f(d)-\theta)^2 p(d,\theta) d\theta.\\
\hat{f}&=&\mathrm{argmin}_f I[f]\\
\end{eqnarray}
This is the form of optimization problem that is solved when training a neural network with squared loss to estimate a parameter $\theta$ from a data $d$. The integral is typically approximated by averaging the squared loss from the training set. The training set is a set of  pairs $(d,\theta)$ that are generated by first sampling from the prior $\theta\leftarrow p(\theta)$ and then simulating the data $d$
from the likelihood $d\leftarrow p(d|\theta)$.

If we assume that the space of neural network functions $f(x,w)$ parameterized by the weights $w$ is sufficiently rich to contain an excellent approximation to $\hat f$, we can simply consider the properties of the optimal function $\hat f$.

We can explicitly show that solution to the optimization problem  $\hat f$ is the posterior mean:
\begin{align}
    \frac{\partial I}{\partial f}&=2\int(\hat{f}(d)-\theta)p(\theta|d)p(d)d\theta=0\\
    &\iff \hat{f}(d)p(d)\int  p(\theta|d) d\theta=p(d)\int \theta p(\theta|d) d\theta\\
    &\iff \hat{f}(d)=\int \theta p(\theta|d) d\theta
\end{align}
Going from the middle to the last line we used the fact that the posterior is normalized and that $p(d)>0$ for any $d$ (otherwise that $d$ would have probability 0 of being in the training set).

\section{B. Neural Network Architecture}
\label{sec:Architectures}

In this appendix we outline the architecture used to train the neural networks whose inputs are the 2D Gaussian density fields and their outputs are the value of the cosmological parameter $A$. The model we use is as follow:

\begin{itemize}
\item Input: Gaussian map with $128\times128$ pixels
\item 2D convolution: kernel=4, stride=2, padding=1 $\longrightarrow~$16 channels $\times~ 64 \times 64$
\item LeakyReLU activation (0.2)
\item 2D convolution: kernel=4, stride=2, padding=1 $\longrightarrow~$32 channels $\times~ 32 \times 32$
\item BatchNorm
\item LeakyReLU activation (0.2)
\item 2D convolution: kernel=4, stride=2, padding=1 $\longrightarrow~$64 channels $\times~ 16 \times 16$
\item BatchNorm
\item LeakyReLU activation (0.2)
\item 2D convolution: kernel=4, stride=2, padding=1 $\longrightarrow~$128 channels $\times~ 8 \times 8$
\item BatchNorm
\item LeakyReLU activation (0.2)
\item 2D convolution: kernel=4, stride=2, padding=1 $\longrightarrow~$256 channels $\times~ 4 \times 4$
\item BatchNorm
\item LeakyReLU activation (0.2)
\item 2D convolution: kernel=4, stride=2, padding=1 $\longrightarrow~$512 channels $\times~ 2 \times 2$
\item BatchNorm
\item LeakyReLU activation (0.2)
\item Flatten $512\times2\times2$ tensor to $2048$ array
\item Fully connected layer $\longrightarrow~$ Output = $A$
\end{itemize}

\section{C. Maximum likelihood with priors versus neural networks}
\label{sec:priors}

In this appendix we attempt to shed light on the structure of the parameter distribution output by the neural network in the case of the Toy model I; upper panel of Fig. \ref{fig:AB_values}.

\begin{figure}
\centering
\includegraphics[width=0.99\textwidth]{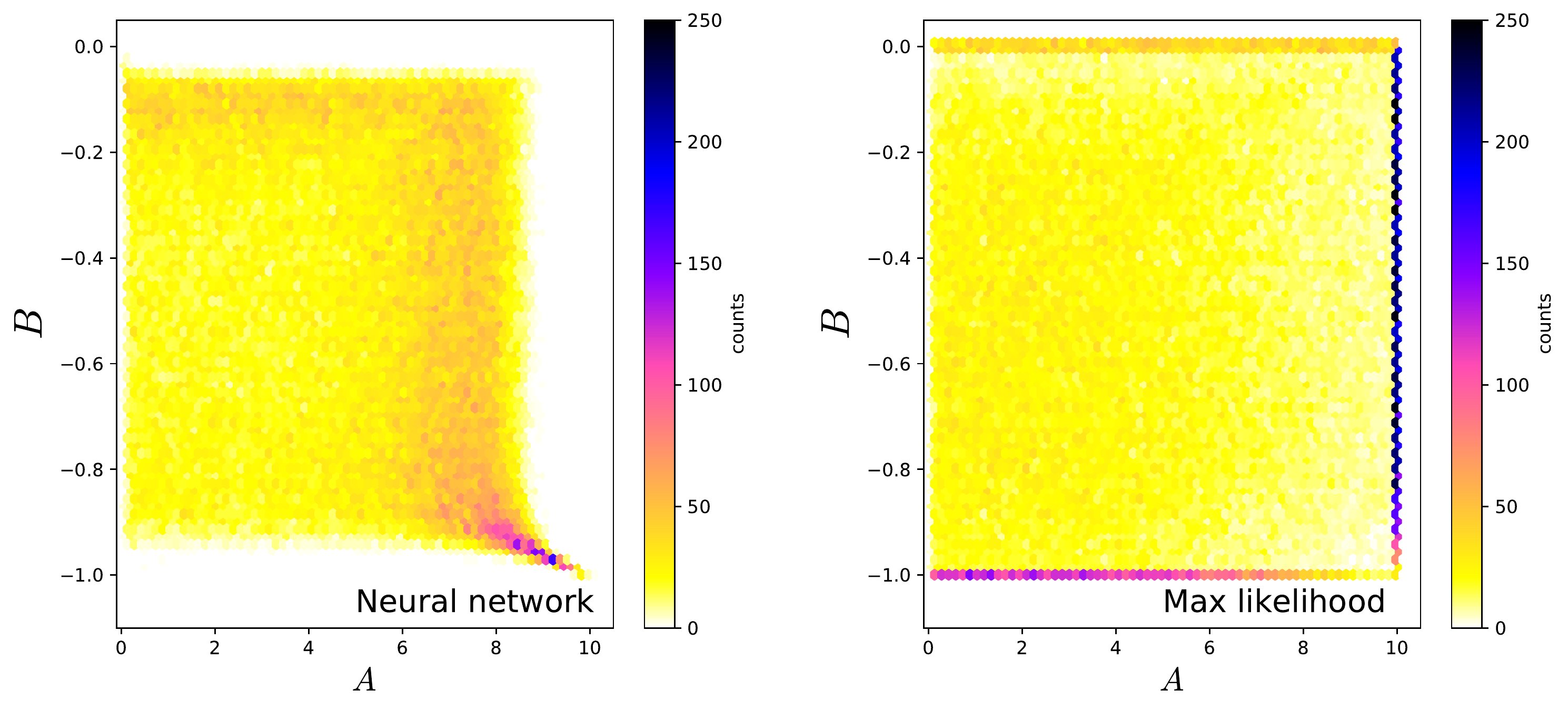}
\caption{Same as Fig. \ref{fig:AB_values} but using the priors $0.1\leq A \leq10$, $-1\leq B \leq 0$ to determine the value of the parameters through the maximum likelihood method. As can be seen, the effects of priors on the maximum likelihood method is to overpopulate the edges; values that will be outside the region constrained by priors are forced to be located on the edges since their likelihood is larger there. We note that results of both methods are different. For instance, the neural network does not seem to make predictions in the region where $A>9$ and $0\leq B \leq 0.9$. The network exhibits a lower variance as estimator, when tested on values across the whole input parameter space, than the maximum likelihood method when priors are taken into account.}
\label{fig:priors}
\end{figure}

We discussed in the main text that the reason why the network never outputs values of $A$ and $B$ outside the range $0.1\leq A \leq 10$, $-1\leq B \leq 0$ is because those are the priors of the distribution it has been trained on. When we computed the values of $A$ and $B$ from the maximum likelihood method we did not take into account the presence of these priors. In Fig. \ref{fig:priors} we show the results when the priors are accounted for when evaluating the likelihood.

While both methods yield similar results, there are some intriguing differences. In the case of the maximum likelihood estimator, the method places a large number of examples in the edges of the parameter distribution. This happens because those points, in the absence of priors, will reside outside the priors region, and the priors move them to the edges, where their likelihood maximizes. This behaviour is however different to the one of the neural network. For instance, the network does not seem to cover the region with $A>9$ and $0\leq B \leq 0.9$. 

\begin{figure}
\centering
\includegraphics[width=0.45\textwidth]{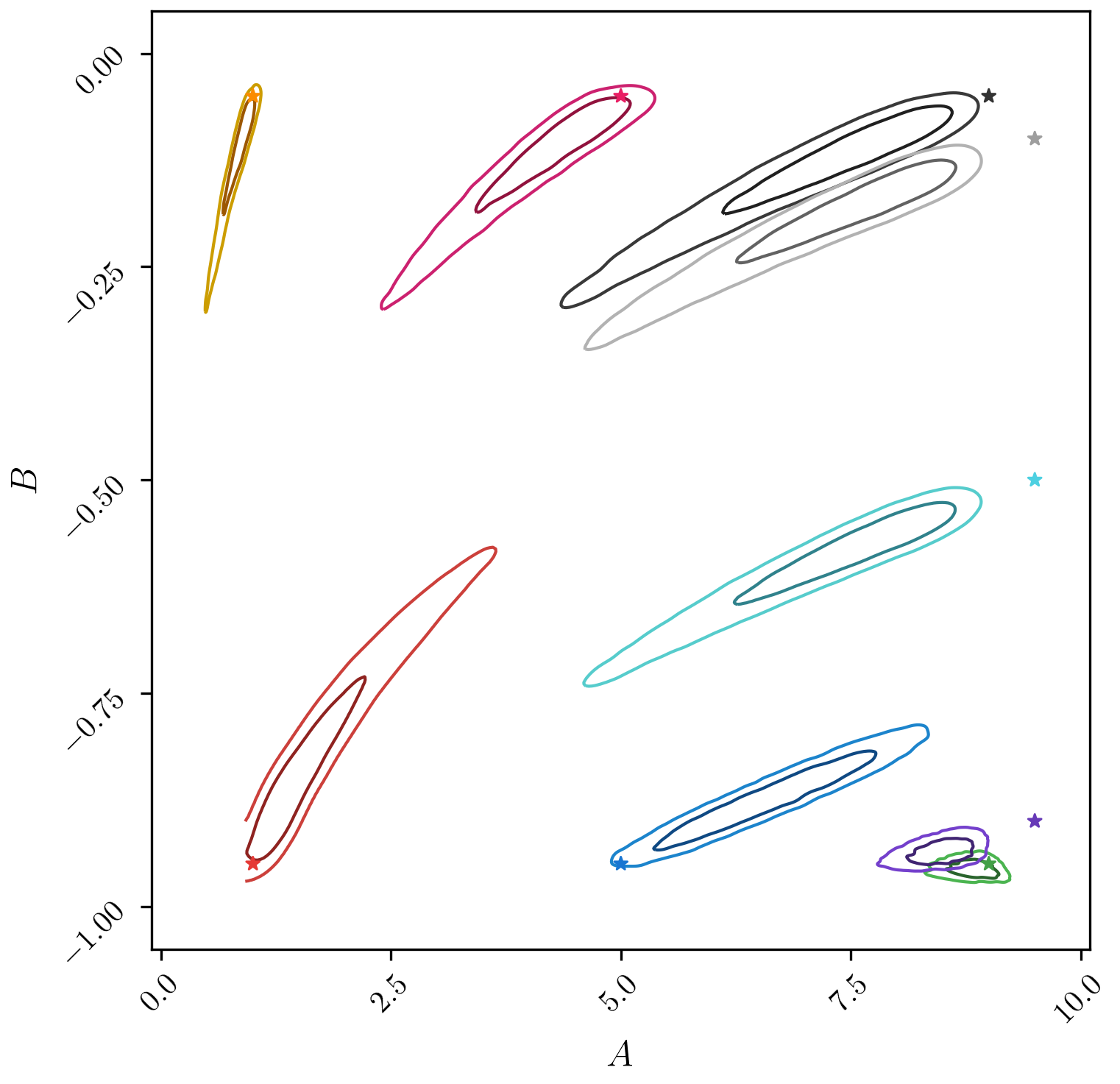}
\caption{We take a point in parameter $(A,B)$ and generate 100,000 power spectra from the AstroNone set with the value of those cosmological parameters. We input those power spectra into a neural network that predicts the values of $A$ and $B$ from measurements of the power spectrum down to $k_{\rm max}=0.05~h{\rm Mpc}^{-1}$. From the predictions of the network, we compute the contours showing the distribution of the data. We have done this exercise for different points in parameter space, $(A,B)$ equal to (9, -0.95), (5,-0.95), (1,-0.95), (1,-0.05), (5,-0.05), (9,-0.05), (9.5, -0.1), (5, -0.1), (9.5, -0.9). This figure shows the results. For values of $A$ smaller than $\simeq9$, the true value in parameter space is within the main confidence intervals of the network predictions, independently of the value of $B$. On the other hand, for values of $A$ greater than $\simeq9$, the network makes most of its predictions far away of the true value. This behavior is distinct from the one of the maximum likelihood with priors (see Fig. \ref{fig:priors}). }
\label{fig:priors2}
\end{figure}

In order to explore this in more detail we have carried out the following exercise. We first take a point in parameter $(A,B)$, and we generate 100,000 power spectra with no baryonic effects (AstroNone set) with the value of those cosmological parameters. We input these power spectra to the network trained for $k_{\rm max}=0.05~h{\rm Mpc}^{-1}$ and compute the distribution of the $A$ and $B$ parameters. We have taken nine different points in parameter space near the boundaries, and show the results in Fig. \ref{fig:priors2}. 

For values of $A$ smaller than $\simeq 9$ we find that the true values lie within the distribution of the neural network predictions, independently of the value of $B$. On the other hand, for values of $A$ larger than 9, the network predicts values of $A$, and to a lesser extent of $B$, that are significantly smaller than the true ones. This seems to be the reason why the neural network does not make predictions for the value of the parameters on that regime. This behaviour can be qualitatively explained taking into account that the network is trying to find an approximation to the posterior mean (see Appendix \ref{sec:posterior_mean}) by integrating over the full support of the posterior. However, a more quantitative interpretation of this effect is beyond the scope of this work, and we will address it in a future work.

\end{appendix}

\vspace{1.0cm}

\bibliography{references}{}
\bibliographystyle{hapj}

\end{document}